\documentclass[12pt,preprint]{aastex}

\def\ltapprox{\raise 2pt \hbox {$<$} \kern-1.1em \lower 5pt \hbox {$\approx$}}
\def\ltsim{\raise 2pt \hbox {$<$} \kern-1.1em \lower 4pt \hbox {$\sim$}}
\def\gtsim{\raise 2pt \hbox {$>$} \kern-1.1em \lower 4pt \hbox {$\sim$}}

\shorttitle{Cosmological effects of powerful AGN outbursts: insights from
MS0735+7421}
\shortauthors{Gitti et al.}

\begin{document}


\title{COSMOLOGICAL EFFECTS OF POWERFUL AGN OUTBURSTS IN GALAXY CLUSTERS:\\
INSIGHTS FROM AN XMM-NEWTON OBSERVATION OF MS0735+7421}


\author{M. Gitti\altaffilmark{1,2},
        B. R. McNamara\altaffilmark{1,3,4}}
\author{P. E. J. Nulsen \altaffilmark{4},
        M. W. Wise \altaffilmark{5}
}

\altaffiltext{1}{Dept. of Physics \& Astronomy, Ohio University,
Clippinger Labs, Athens, OH 45701 - USA}
\altaffiltext{2}{INAF - Osservatorio Astronomico di Bologna,
via Ranzani 1, I-40127 Bologna  - Italy}
\altaffiltext{3}{Dept. of Physics \& Astronomy, University of Waterloo,
200 University Avenue West, Waterloo, Ontario - Canada N2L 2G1}
\altaffiltext{4}{Harvard-Smithsonian Center for Astrophysics,
60 Garden Street, Cambridge, MA 02138 - USA}
\altaffiltext{5}{Astronomical Institute, University of Amsterdam, Kruislaan 
403, 1098 SJ Amsterdam - The Netherlands}




\begin{abstract}
We report on the results of an analysis of \textit{XMM-Newton} observations of
MS0735+7421, the galaxy cluster which hosts the most energetic AGN outburst
currently known.
The previous \textit{Chandra} image shows twin giant X-ray cavities 
($\sim$ 200 kpc diameter)
filled with radio emission and surrounded by a weak shock front.
\textit{XMM} data are consistent with these findings.
The total energy in cavities and shock
($\sim 6 \times 10^{61}$ erg) is enough to quench the 
cooling flow and, since most of the energy is 
deposited outside the cooling region ($\sim$ 100 kpc), to 
heat the gas within 1 Mpc by $\sim$ 1/4 keV per particle.
The cluster exhibits an upward departure (factor $\sim$2) from the mean 
$L$-$T$ relation.
The boost in emissivity produced by the ICM compression in the bright shells 
due to the cavity expansion may contribute to explain the high luminosity
and high central gas mass fraction that we measure. 
The scaled temperature and metallicity profiles are in 
general agreement with those observed in relaxed clusters.
Also, the quantities we measure are consistent with the observed
$M$-$T$ relation.  
We conclude that violent outbursts such as the one in MS0735+7421 do not 
cause dramatic instantaneous departures from cluster scaling relations 
(other than the $L$-$T$ relation).   
However, if they are relatively common they may play a role in 
creating the global cluster properties.
\end{abstract}




\section{Introduction}
\label{intro.sec}

The current generation of X-ray satellites, \textit{Chandra} and 
\textit{XMM-Newton}, has shown that the physics of the intra-cluster medium 
(ICM) is complex and needs to be regulated by additional,
non-gravitational processes beyond simple gravity and gas dynamics
considered in the standard Cold Dark Matter cosmological scenario 
(White \& Rees 1978).
In particular, our understanding of cooling flow systems has radically
changed.
Albeit confirming the existence of short cooling times, high densities and low
temperatures in the cluster cores, the arrival of new high-resolution X-ray
data has shown the lack of the emission lines expected from gas cooling
below 1-2 keV and has reduced by about one order of magnitude the
new spectroscopically--derived mass deposition rates
(e.g. David et al. 2001, Johnstone et al. 2002, Peterson et al. 2003
and references therein).
The most plausible solution to this so-called ``cooling flow problem'' is
that some form of additional heating which
balances the cooling must be acting in the ICM.
Among the many proposed heating mechanisms\footnote{
Proposed heating
mechanisms include electron thermal conduction from the outer
regions of clusters (Tucker \& Rosner 1983; Voigt et al. 2002;
Fabian et al. 2002; Zakamska \& Narayan 2003), continuous
subcluster merging (Markevitch et al. 2001), contribution of the
gravitational potential of the cluster core (Fabian 2003),
feedback from intra-cluster supernovae (Domainko et al. 2004),
etc.
}, one of the best candidates for supplying the energy is feedback from the
Active Galactic Nucleus (AGN) hosted by the central galaxy of the
cluster (e.g., Rosner \& Tucker 1989; Tabor \& Binney 1993;
Churazov et al. 2001; Br{\"u}ggen \& Kaiser 2001; Kaiser \& Binney 2003;
Ruszkowski \& Begelman 2002; Brighenti \& Mathews 2003, Omma et al. 2004).

The possibility of AGN heating has recently become the
leading idea to solve the cooling flow problem thanks to the
discovery of X-ray cavities in the ICM on scales often approximately 
coincident with the lobes of extended radio emission 
(e.g., Hydra A: McNamara et al. 2000, David et al. 2001; 
Perseus: B\"ohringer et al. 1993, Churazov et al. 2000,
Fabian et al. 2000; A2052: Blanton et al. 2001, 2003; A2597:
McNamara et al. 2001, Pollack et al. 2005, Clarke et al. 2005; 
RBS797: Schindler et al. 2001, Gitti et al. 2006; A4059: Heinz et al. 2002).
This indicates that the radio lobes fed by AGNs have
displaced the X-ray emitting gas, creating cavities in the ICM.
These outbursts can be recurrent, as it has been observed in
some cases. For example, RBS797 shows evidence of a restarted AGN
activity with precessing jets pointing to a different direction with respect
to the one of the outer radio lobes filling the cavities
(Gitti et al. 2006).
The heating by AGN is thought to occur through the dissipation of the 
cavity enthalpy and through shocks driven by the bubbles 
inflated by the AGN.
Systematic studies of a sample of X-ray cavities show that their 
enthalpies\footnote{
The enthalpy (free energy) of the cavities is estimated
from the total AGN power as $H = \frac{\gamma}{\gamma -1} PV$,
where the $PV$ work done by the jet as it inflates the cavity is 
determined by measuring the cavity size and its surrounding pressure,
and the adiabatic index $\gamma$ is related to the internal composition
of the cavities, still unknown.
In the case of a relativistic plasma $\gamma$ = 4/3 and $H=4PV$.
}
lie between $10^{55}-10^{61}$ erg and scales in
proportion to the cooling X-ray luminosity and the radio power of
the host system.
In more than a half of the sample the energy input from the
central radio source is currently sufficient to balance cooling
(B\^{\i}rzan et al. 2004, Rafferty et al. 2006).
The trend between X-ray luminosity and bubble mechanical
luminosity, together with the existence of short central cooling
time, suggests that the AGN is fueled by a cooling flow that is itself
regulated by feedback from the AGN.
The basic idea of this AGN-cooling flow scenario is that a
self-regulated equilibrium may be achieved, in which the energy
input from the AGN balances the radiative losses of the ICM over
the lifetime of the cluster.

The recent discovery of giant cavities and associated
large-scale shocks in three system (MS0735+7241: McNamara et al. 2005,
Hercules A: Nulsen et al. 2005a, Hydra A: Nulsen et al. 2005b)
has shown that AGN outbursts can not only
affect the central regions, but also have an impact on
cluster-wide scales.
In particular, the supercavities discovered in MS0735+7421 have a
diameter of about 200 kpc each and a weak cocoon shock surrounding them has
been detected. This large-scale outburst is the most powerful known so far:
it releases upward of $10^{61}$ erg into the ICM, heating the gas beyond the
cooling region (McNamara et al. 2005).
In this paper we want to investigate the significant consequences that
this new development has for several fundamental problems in astrophysics.
One important problem addressed by giant cavities is the
so-called cluster "pre-heating", which is manifested in the
steepening of the observed luminosity vs. temperature relation for
clusters with respect to theoretical predictions that include
gravity alone (e.g., Markevitch 1998).
Some extra non-gravitational energy is required to
explain such a steepening (e.g., Wu et al. 2000, Voit 2005),
and one possibility is that it is supplied by AGN outburst.
The additional non-gravitational heating supplied by AGN
could also induce the suppression of the gas cooling in massive
galaxies required to explain the exponential turnover at the
bright end of the luminosity function of galaxies (Benson et al. 2003).
This would indicate a common solution for the two major heating problems
associated with the ICM: those of cooling flow and galaxy
formation.
Some cavities live longer than the time it takes to cross their own
diameters, suggesting they are in pressure balance 
with the surrounding medium.
Since the work required to inflate the radio lobes generally
exceeds the nonthermal minimum energy of the radio emission
(e.g., De Young 2006 and references therein), 
in the hypothesis that the assumptions adopted into the equipartition
calculation are correct, some additional component besides the 
relativistic plasma may fill the cavity and
contribute to support them internally.
We want to investigate what is currently providing the necessary pressure
support to sustain the cavities.
Finally, in a more general view, we want to investigate what
is the potential impact that such powerful outbursts have on
the global properties of the ICM.
Understanding well the ICM physics is essential to use galaxy
clusters as high-precision cosmological tools. 

We address these problems by studying the X-ray properties of the most
energetic outburst known in a galaxy cluster.
MS0735$+$7421 (hereafter MS0735) is at a redshift of 0.216.
With a Hubble constant of $H_0 = 70 \mbox{ km s}^{-1} \mbox{ Mpc}^{-1}$, and
$\Omega_M = 1-\Omega_{\Lambda} = 0.3$, the luminosity distance is 1069 Mpc
and the angular scale is 3.5 kpc per arcsec.


\section{Observation and data preparation}
\label{observation.sec}

MS0735 was observed by \textit{XMM--Newton} in April 2005 during
revolution 0973.
In this paper only data from the European Photon Imaging Camera (EPIC)
are analyzed and discussed, while those from the Reflection Grating 
Spectrometer (RGS) will be presented in a subsequent paper. 
The MOS and pn detectors were both operated in Full Frame Mode
with the THIN filter, for an exposure time of 70.6 ks for MOS and
60.7 ks for pn. 
We use the SASv6.5.0 processing tasks
\textit{emchain} and \textit{epchain} to generate calibrated event
files from raw data. Throughout this analysis single pixel events
for the pn data (PATTERN 0) are selected, while for the MOS data
sets the PATTERNs 0-12 are used. The removal of bright pixels and
hot columns is done in a conservative way applying the expression
(FLAG==0). To reject the soft proton flares we accumulate the
light curve in the [10-12] keV band for MOS and [12-14] keV band
for pn, where the emission is dominated by the particle--induced
background, and exclude all the intervals of exposure time having
a count rate higher than a certain threshold value (the chosen
threshold values are 0.4 cps for MOS and 0.7 cps
for pn). The remaining exposure times after cleaning are 50.4 ks
for MOS1, 49.5 ks for MOS2 and 42.0 ks for pn. Starting from the
output of the SAS detection source task, we make a visual
selection on a wide energy band MOS \& pn image of point sources
in the field of view. Events from these regions are excluded
directly from each event list. The source and background events
are corrected for vignetting using the weighted method described
in Arnaud et al. (2001), the weight coefficients being tabulated
in the event list with the SAS task \textit{evigweight}. This
allows us to use the on-axis response matrices and effective
areas.

Unless otherwise stated, the reported errors are at 90\% confidence level.


\subsection{Background treatment}
\label{background.sec}

The background estimates are obtained using a blank-sky
observation consisting of several high-latitude pointings with
sources removed (Lumb et al. 2002). The blank-sky background
events are selected using the same selection criteria (such as
PATTERN, FLAG, etc.), intensity filter (for flare rejection) and
point source removal used for the observation events. This yields
final exposure times for the blank fields of 365 ks for MOS1,
350 ks for MOS2 and 294 ks for pn. Since the cosmic ray
induced background might change slightly with time, we compute the
ratio of the total count rates in the high energy band ([10-12]
keV for MOS and [12-14] keV for pn). The obtained normalization
factors (1.266, 1.303, 1.283 for MOS1, MOS2 and pn, respectively)
are then used to renormalize the blank field data. The blank-sky
background files are recast in order to have the same sky
coordinates as MS0735.

The usual approach to perform the background subtraction is
described in full detail in Arnaud et al. (2002). This procedure
consists of two steps. In a first step, for each product extracted
from the observation event list an equivalent product is
extracted from the corresponding blank-field file and then
subtracted from it. This allows us to remove the particle
background. However, if the background in the observation region
is different from the average background in blank field data, this
step could leave a residual background component. The residual
background component is estimated by using blank-field-subtracted
data in a region free of cluster emission (in particular we
consider an annulus lying between 8 and 10 arcmin) and then
subtracted in a second step from each MOS and pn product. In our
case the residual is negative in the energy band adopted for the
spatial analysis ([0.4-2] keV, see Sect. \ref{brightness.sec}).
The residual count rate summed over the three detectors is $-0.07$
counts/s, which represents $\sim 10 \%$ of the total background
count rate in this energy band.

The fluorescent emission lines\footnote{
Al and Si lines at 1.5 and 1.7 keV, respectively, in the MOS data, 
and Ni, Cu and Zn lines around 8 keV in the pn data.} 
excited by the energetic
charged particles that pass through the detector exhibit spatial
variation over the detector. This effect would compromise the
reliability of the second step of the background subtraction for
spectra, as the resulting total background spectrum (sum of the
particle background spectrum, estimated in the
blank-field-subtracted cluster region, and the residual background
component, estimated in the blank-field-subtracted outer annular
region) shows fluorescence lines which are shifted by a few energy
channels with respect to the MS0735 spectrum. Therefore the
residual background component is neglected for the purpose of the
spectral analysis.


\section{Morphological analysis}
\label{morphology.sec}

\begin{figure}
\epsscale{.80}
\plotone{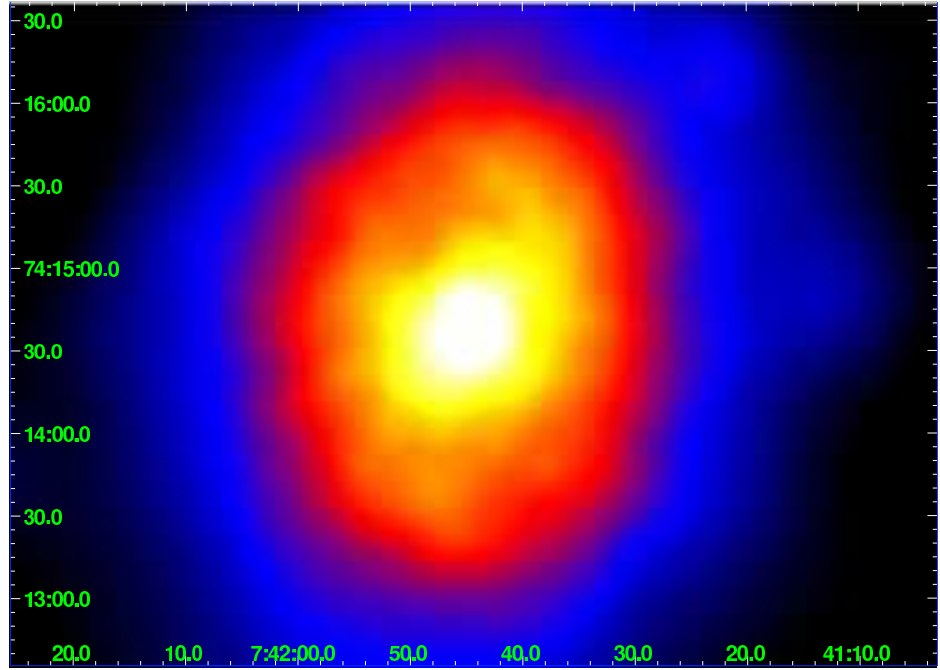}
\caption{
MOS1 image of MS0735 in the [0.3-10] keV energy band.
The image is corrected for vignetting and
exposure and is adaptively smoothed (signal-to-noise ratio = 20).
\label{smooth.fig}}
\end{figure}

The adaptively smoothed, exposure corrected MOS1 count rate image
in the [0.3-10] keV energy band is presented in Fig. \ref{smooth.fig}.
The smoothed image is obtained from the raw image corrected for the
exposure map (that accounts for spatial quantum efficiency,
mirror vignetting and field of view) by running the task
\textit{asmooth} set to a desired signal-to-noise ratio of 20.
Regions exposed with less than 10\% of the total exposure are not considered.

The inner part of the cluster shows the high surface brightness characteristic
of a cooling flow.
We notice a sharp central surface brightness peak at a position
$07^{\rm h} 41^{\rm m} 44^{\rm s}.06 \, +74^{\circ} 14' 38''.62$ (J2000).
At small radii, 
two strong depressions having a diameter of about 150 kpc
(see Sect. \ref{brightness.sec}) are visible on opposite side of
the cluster center in the NE-SW direction.
The cavities are surrounded by a bright X-ray emission
of elliptical shape, in the form expected by a radio cocoon.
The elliptical discontinuity in the X-ray surface brightness appears more
evident in the \textit{Chandra} data and has been interpreted as a weak shock
(McNamara et al. 2005).
Beyond the cavity region the cluster maintains a slightly elliptical morphology
up to large radii, showing hints of structures that could be the results of
past radio activity that created older cavities.

\subsection{Surface brightness profile}
\label{brightness.sec}

We consider different sectors to the north and south
(N sector: between 10$^{\circ}$ west and 60$^{\circ}$ east of north;
S sector:  between 15$^{\circ}$ east and 60$^{\circ}$ west of south)
and to the east and west
(E sector: between 30$^{\circ}$ north and 75$^{\circ}$ south of east;
W sector:  between 30$^{\circ}$ south and 80$^{\circ}$ north of west)
in order to include and exclude the cavity regions, respectively.
We also consider a full $360^{\circ}$ sector in which we mask the cavities
(hereafter undisturbed cluster).
For each sector, we compute a background-subtracted, vignetting-corrected,
radial surface brightness profile in the [0.4-2] keV energy band for each
camera separately. For the pn data, we generate a list of out-of-time
events\footnote{Out-of-time events are caused by photons which
arrive while the CCD is being read out, and are visible in an
uncorrected image as a bright streak smeared out in RAWY.}
(hereafter OoT) to be treated as an additional background
component. The effect of OoT in the current observing mode (Full
Frame) is 6.3\%. The OoT event list is processed in a similar way
as done for the pn observation event file. The profiles for the
three detectors are then added into a single profile, binned such
that at least a signal-to-noise ratio of 3 is reached.

\begin{figure}
\epsscale{.80}
\plotone{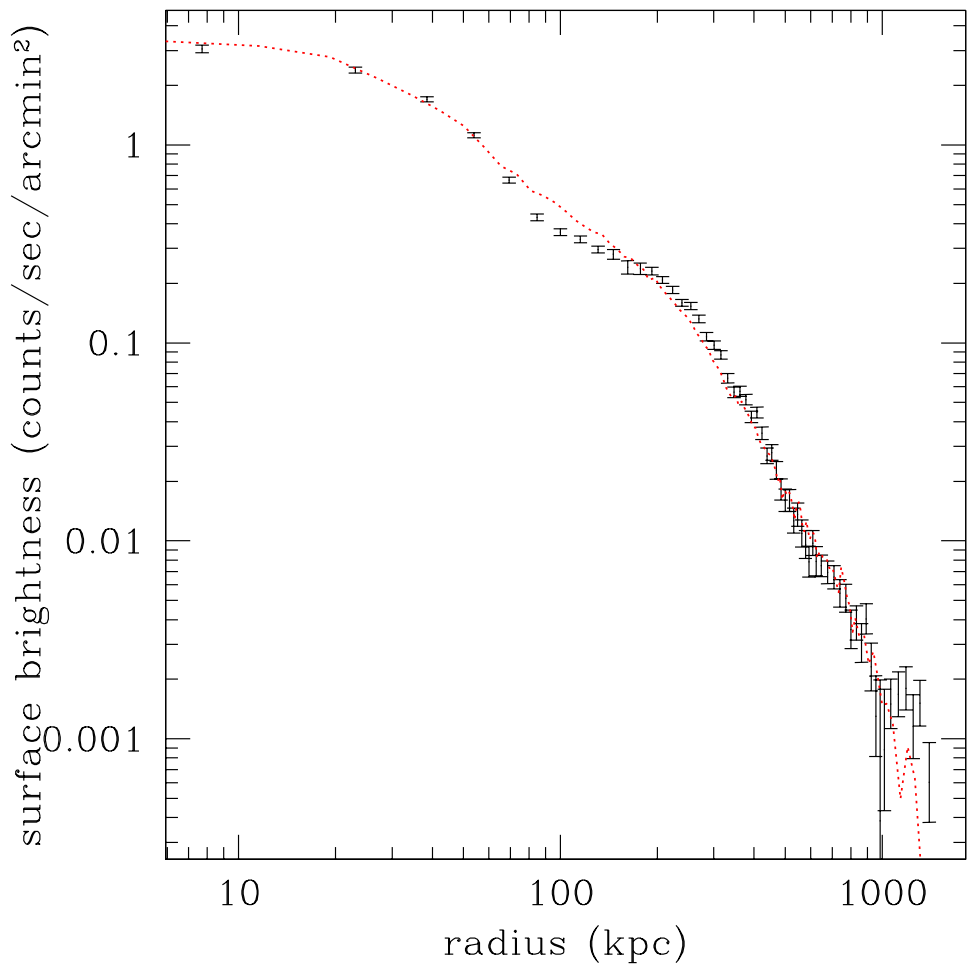}
\caption{
Background subtracted, azimuthally-averaged radial surface brightness profile
of the N sector data in the [0.4-2] keV range.
The dotted red line shows the profile in full $360^{\circ}$ sector with
masked cavities (undisturbed cluster).
A deficit of emission in the N sector between radii of $\sim$ 50-200 kpc
is visible.
\label{sbprofile.fig}}
\end{figure}

The cluster emission is detected up to 1.3 Mpc ($\sim 6$ arcmin).
In Fig. \ref{sbprofile.fig} we show the X-ray surface brightness
profile for the sector containing the northern cavity compared
from that of the undisturbed cluster. We note that the data of the
undisturbed cluster appear regular, while those for the N sector
show a clear deficit of emission between radii $\sim 50$ and
$\sim$ 200 kpc relative to the other directions. The S sector
profile shows a behavior similar to the N one, although the
depression is less pronounced, whereas the E and W sectors appear
regular.

\begin{figure}
\epsscale{.80}
\plotone{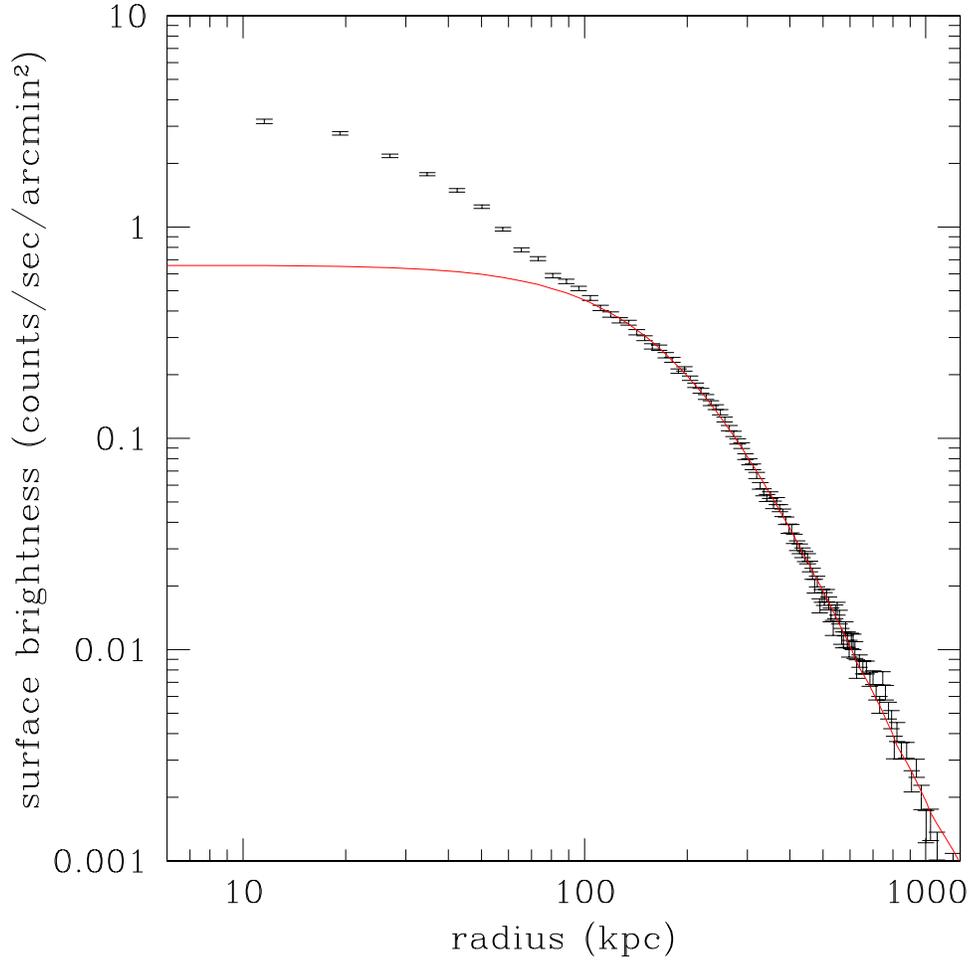}
\caption{
Background subtracted, azimuthally-averaged radial surface brightness profile
in the [0.4-2] keV range.
The best fit $\beta$-model
fitted over the 100 - 1150 kpc region is over-plotted as a solid red line.
In the central region, the observed surface brightness profile shows a 
strong excess as compared to the extrapolation of the model.
\label{1betafit.fig}}
\end{figure}

The surface brightness profile of the undisturbed cluster is fitted in the
CIAO tool \textit{Sherpa} with various parametric models, which are
convolved with the \textit{XMM} point spread function (PSF).
The overall PSF is obtained by adding the PSF of each camera (Ghizzardi 2001),
estimated at an energy of 1.5 keV and weighted by the respective cluster
count rate in the [0.4-2] keV energy band.
A single $\beta$-model (Cavaliere \& Fusco Femiano 1976) is not a good
description of the entire profile: a fit to the outer regions
shows a strong excess in the center as compared to the model
(see Fig. \ref{1betafit.fig}).
The centrally peaked emission is a strong indication of a cooling flow in this
cluster.
We find that for 100 kpc \ltsim r \ltsim 1150 kpc the data can be described
($\chi^2_{\rm red} \sim 1.65$ for 87 d.o.f.) by a
$\beta$-model with a core radius $r_{\rm c}=195 \pm 4$ kpc and a slope
parameter $\beta=0.77 \pm 0.01$ (3 $\sigma$ confidence level).
The single $\beta$-model functional form is a convenient representation
of the gas density profile in the outer regions, which is used as a tracer
for the potential. This best fit model is thus used in the following
to estimate the cluster gas and total mass profiles
(see Sect. \ref{mass.sec}).

We also consider a double isothermal $\beta$-model and find that it can
account for the entire profile, when the very inner and outer regions are
excluded:
for 10 kpc \ltsim r \ltsim 1150 kpc the best fit parameters are
$r_{\rm c1}=200 \pm 4$ kpc, $\beta_1=0.77 \pm 0.01$,
$r_{\rm c2}=156 \pm 8$ kpc, $\beta_2=4.67 \pm 0.56$;
$\chi^2_{\rm red} \sim 1.87$ for 96 d.o.f.
By assuming a common $\beta$ value we find:
$r_{\rm c1}=215 \pm 4$ kpc, $r_{\rm c2}=30 \pm 2$ kpc,
$\beta=0.79 \pm 0.01$;
$\chi^2_{\rm red} \sim 2.08$ for 97 d.o.f.

\section{Temperature map}
\label{tmap.sec}

\begin{figure}
\epsscale{.80}
\plotone{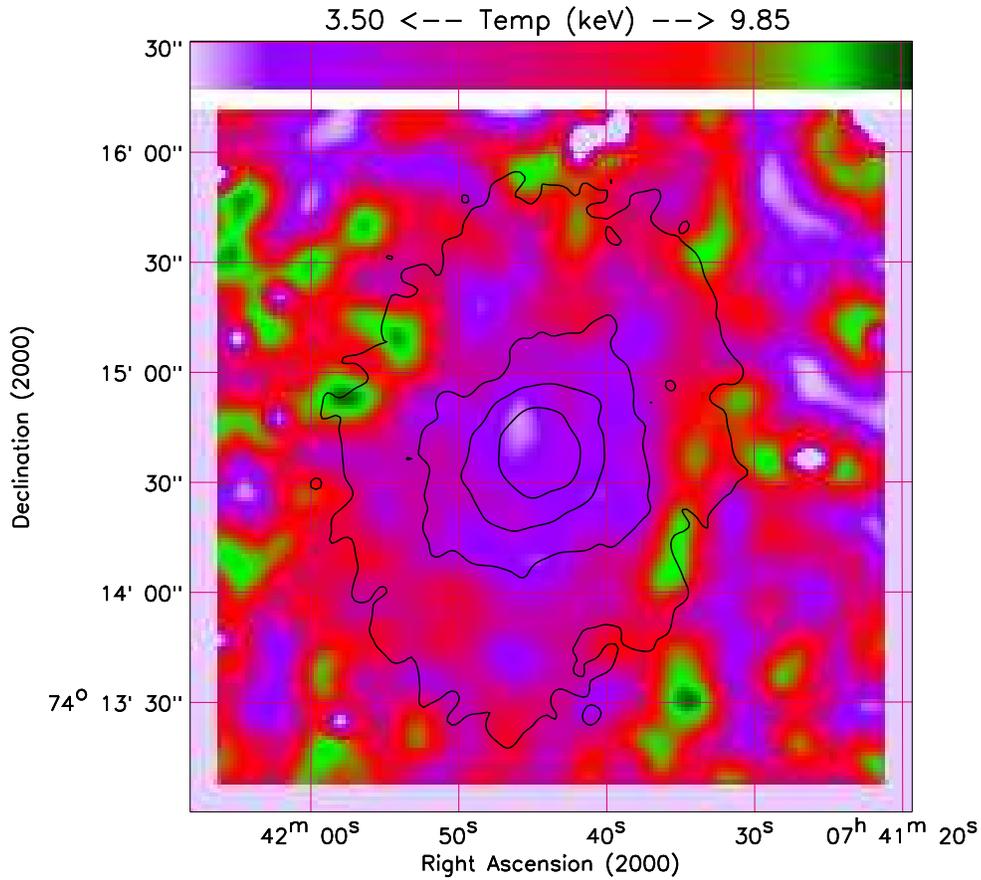}
\caption{Temperature map obtained by using 4 X-ray colors ([0.3-1.0], 
[1.0-2.0], [2.0-4.5], [4.5-8.0] keV) and estimating the expected count 
rate with XSPEC for a thermal {\ttfamily mekal} model,  
with fixed absorbing column $N_{\rm H} = 3.49 \times 10^{20} \mbox{ atom cm}^{-2}$
and metallicity $Z=0.4 \, Z_{\sun}$.
The metallicity gradient in the central region 
(see Fig. \ref{profilomet-proj.fig}) does not affect the map significantly.
The map has a pixel size of 1.1 arcsec and is smoothed with a Gaussian of
$\sigma = 16.5$ arcsec.
Superposed are the X-ray contours. The features outside the last contours are
not significant, as they are mainly due to noise fluctuations.
\label{tmap.fig}}
\end{figure}

The temperature image of the cluster central region shown in Fig.
\ref{tmap.fig} is built from X-ray colors. Specifically, we extract
mosaiced MOS images in four different energy bands ([0.3-1.0] keV, [1.0-2.0]
keV, [2.0-4.5] keV and [4.5-8.0] keV), subtract the background and divide
the resulting images by the exposure maps.
A temperature in each pixel of the map is
obtained by fitting values in each pixel of these images with a thermal
plasma, fixing $N_H$ to the Galactic value (Dickey \& Lockman 1990)
and the element abundance to 
0.4 solar (see Sect. \ref{global.sec}).
Besides the evidence that the very central region
is cooler than the surrounding medium, we do not notice any particular
structure in the temperature distribution.

The regularity of the temperature distribution points to a relaxed dynamical
state of the cluster, thus excluding the presence of an ongoing merger.
Since cluster merging can cause strong deviations from the assumption of an
equilibrium configuration, this allows us to derive a good estimate of the
cluster mass (see Sect. \ref{mass.sec}).

\section{Spectral analysis}
\label{spectral.sec}

Throughout the analysis, a single spectrum is extracted for each region of
interest and is then regrouped to give at least 25 counts in each bin.
The data are modeled using the XSPEC code, version 11.3.0.
Unless otherwise stated, the relative normalizations
of the MOS and pn spectra are left free when fitted simultaneously.
We use the following response matrices:
{\ttfamily m1{\_}534{\_}im{\_}pall{\_}v1.2.rmf} (MOS1),
{\ttfamily m2{\_}534{\_}im{\_}pall{\_}v1.2.rmf} (MOS2),\\
{\ttfamily epn{\_}ff20{\_}sY9{\_}v6.7.rmf} (pn).


\subsection{Global spectrum}
\label{global.sec}

For each instrument, a global spectrum is extracted from all events lying
within 6 arcmin from the cluster emission peak,
which corresponds to the outermost radius determined from the
morphological analysis (Sect. \ref{brightness.sec}). 
We test in detail the consistency between the three cameras
by fitting separately these spectra with an absorbed {\ttfamily mekal}
model with
the redshift fixed at z=0.216 and the absorbing column fixed at the galactic
value ($N_{\rm H} = 3.49 \times 10^{20} \mbox{ atom cm}^{-2}$,
Dickey \& Lockman 1990) and studying the effect of imposing various high and
low-energy cutoffs. 
We find good agreement between the three cameras in the [0.4-10.0] keV
energy range
($kT = 4.64^{+0.17}_{-0.17}$ keV for MOS1, $4.41^{+0.16}_{-0.16}$ keV
for MOS2, $4.35^{+0.13}_{-0.12}$ keV for pn).

The combined MOS+pn global temperature 
(in keV) and metallicity 
(as a fraction of the solar value, Anders \& Grevesse 1989)
derived from the best fit ($\chi^2_{\rm red} \sim 1.18$ for 1947 d.o.f.)
are respectively: $kT = 4.43^{+0.09}_{-0.08}$ keV, 
$Z = 0.35^{+0.03}_{-0.03} \, \rm{Z}_{\sun}$. 
The unabsorbed luminosities in this model
(estimated from the average of the fluxes measured by the three
cameras after fixing $N_{\rm H}=0$) 
in the X-ray ([2.0-10.0] keV) and bolometric band are
respectively: $L_X = 4.61 \pm 0.04 \times 10^{44} \mbox{ erg
s}^{-1}$, $L_{\mbox{bol}}= 1.05 \pm 0.01 \times 10^{45} \mbox{ erg
s}^{-1}$, where the errors are given as half the difference
between the maximum and the minimum value.


\subsection{Projected radial profiles: temperature and metallicity}
\label{radial.sec}

We produce projected radial temperature and metallicity profiles
by extracting spectra in circular annuli centered on the peak of the X--ray
emission. The annular regions are detailed in Table
\ref{profile_proj.tab}. The data from the three cameras are
modelled simultaneously using a simple, single-temperature model
({\ttfamily mekal} plasma emission code in XSPEC) with the
absorbing column density fixed at the nominal Galactic value. The
free parameters in this model are the temperature $kT$,
metallicity $Z$ (measured relative to the solar values, with the
various elements assumed to be present in their solar ratios,
Anders \& Grevesse 1989)
and normalization (emission measure). The best-fitting parameter
values and 90\% confidence levels derived from the fits to the
annular spectra are summarized in Table \ref{profile_proj.tab}.

\begin{figure}
\epsscale{.80}
\plotone{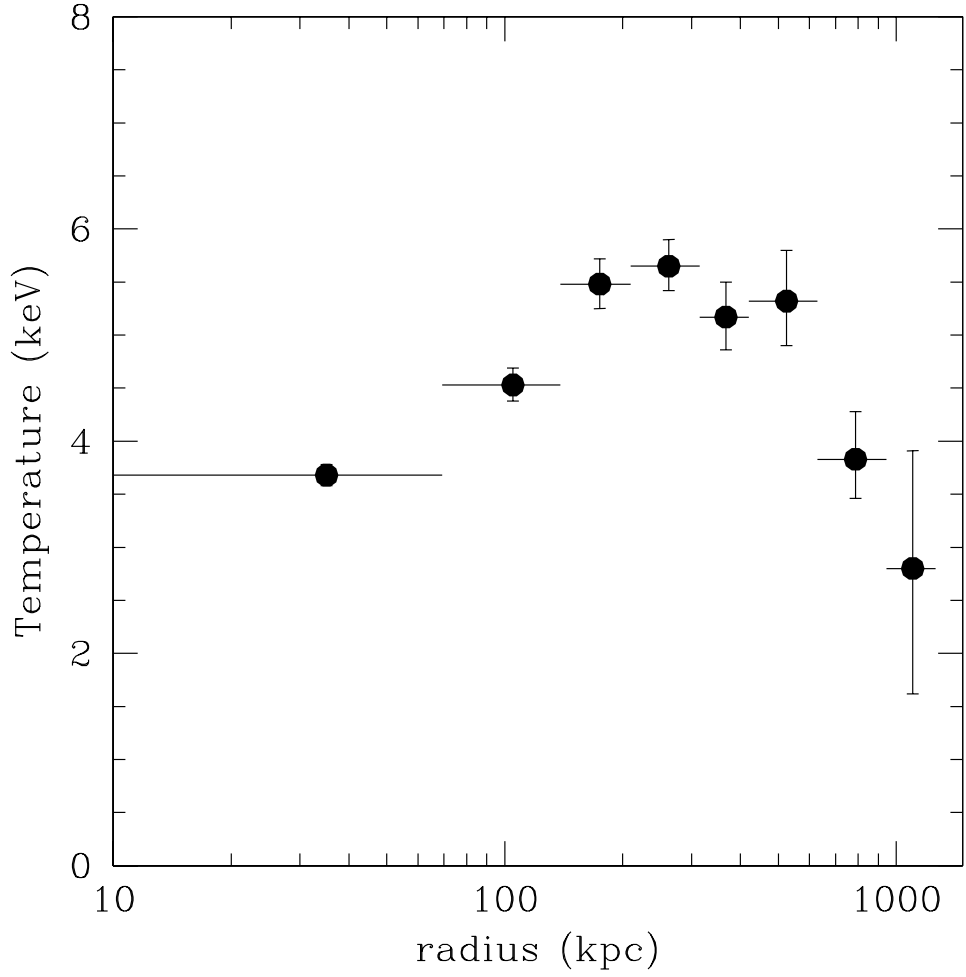} 
\caption{Projected X-ray gas temperature profile measured in the 
[0.4-10.0] keV energy range. 
\label{profilot-proj.fig}}
\end{figure}

The projected temperature profile determined with this model is
shown in Fig. \ref{profilot-proj.fig}. It shows a rise from a mean
value of $\sim 3.7$ keV within 70 kpc to $\sim 5.4$ keV over the
150-630 kpc region, then it declines to a value $\sim 2.8$
keV in the outskirts of the cluster (up to 1.3 Mpc).

\begin{figure}
\epsscale{.80}
\plotone{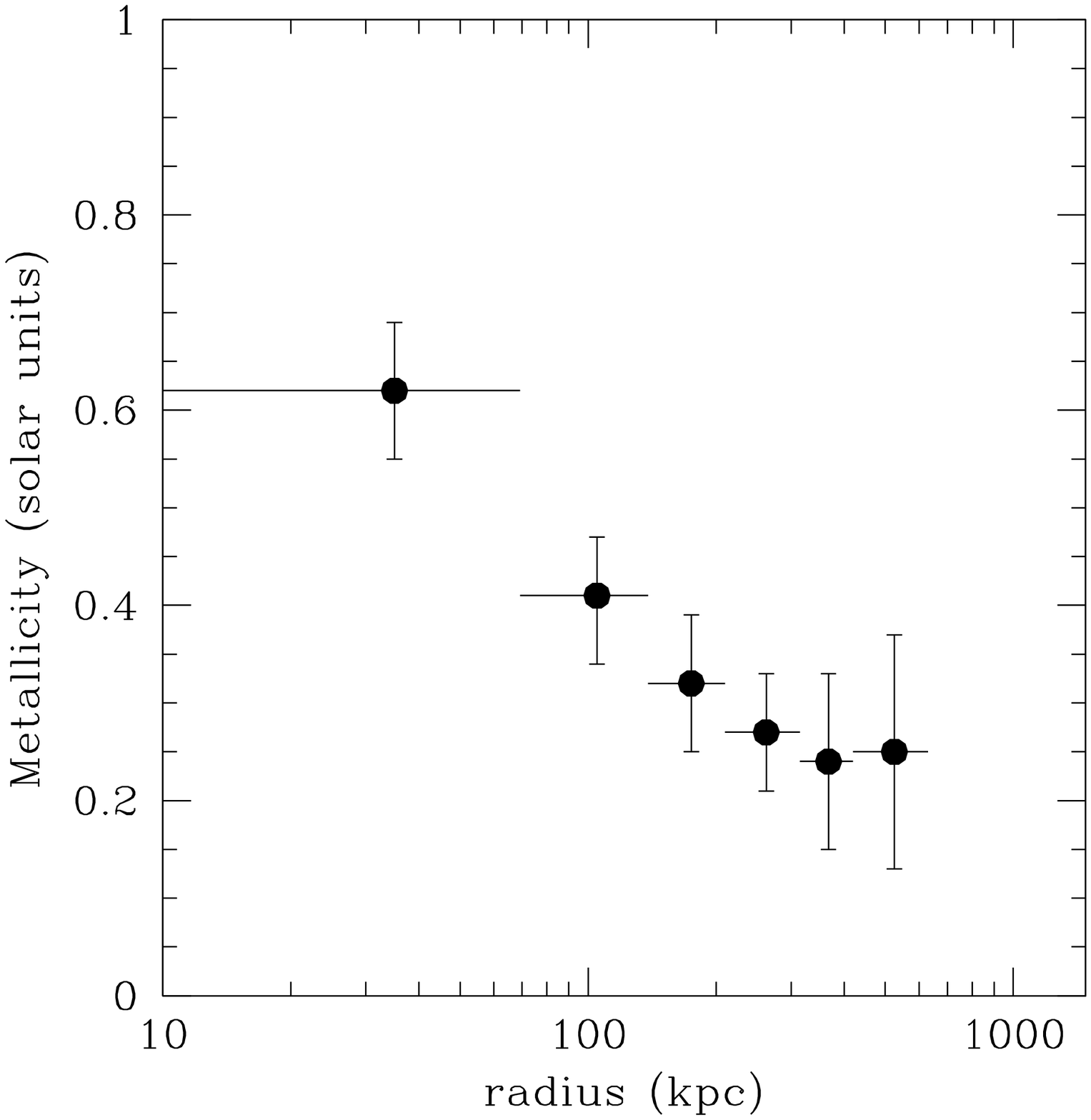} 
\caption{Projected X-ray gas metallicity profile measured 
in the [0.4-10.0] keV energy range. 
The last two bins have been excluded due to poor photon statistics.} 
\label{profilomet-proj.fig}
\end{figure}

The metallicity profile is shown in Fig.
\ref{profilomet-proj.fig}: a gradient is visible towards the
central region, the metallicity increasing from $Z \sim 0.27 \,
 \rm{ Z}_{\sun}$  over the 150-630 kpc region to $Z \sim 0.62 \,
 \rm{ Z}_{\sun}$ inside the central 70 kpc. 
Due to the poor photon statistics that does not allow us to derive 
accurate measurements, we exclude the last two bins. 

We also perform the spectral fitting by leaving the absorbing column density
as a free parameter and find little variation in the results.

\subsection{Deprojection analysis: temperature, density, pressure and
entropy profiles}
\label{deproj.sec}

Because of projection effects, the spectral properties at any
point in the cluster are the emission-weighted superposition of
radiation originating at all points along the line of sight
through the cluster. To correct for the effect of this projection,
we perform a deprojection analysis on the same annular spectra
used in Sect. \ref{radial.sec} by adopting the XSPEC {\ttfamily
projct} model. Under the assumption of ellipsoidal (in our
specific case, spherical) shells of emission, this model
calculates the geometric weighting factor according to which the
emission is redistributed amongst the projected annuli.

\begin{figure}
\epsscale{.80}
\plotone{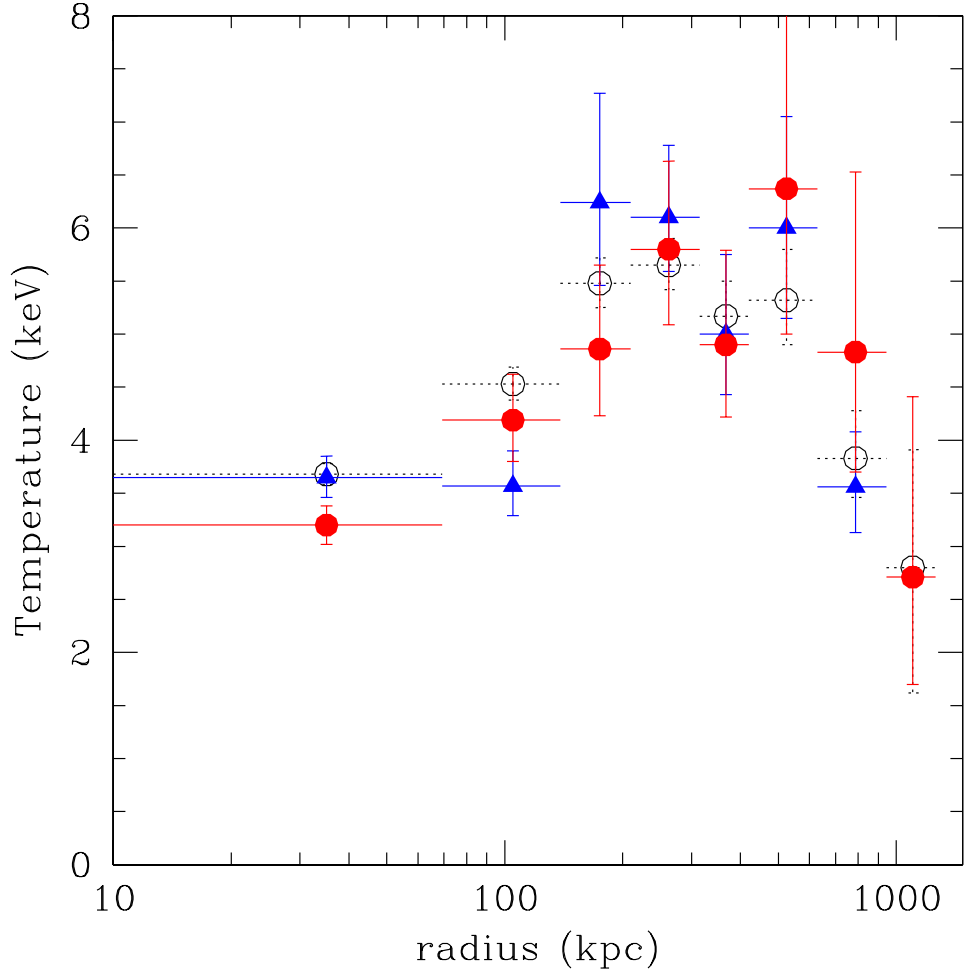}
\caption{
Deprojected X-ray gas temperature profile measured from MOS1+MOS2 
(blue triangles) and pn (red full circles) spectra. 
For comparison, the projected profile is also shown (black open circles).
\label{profilot-deproj.fig}}
\end{figure}

\begin{figure}
\epsscale{.80}
\plotone{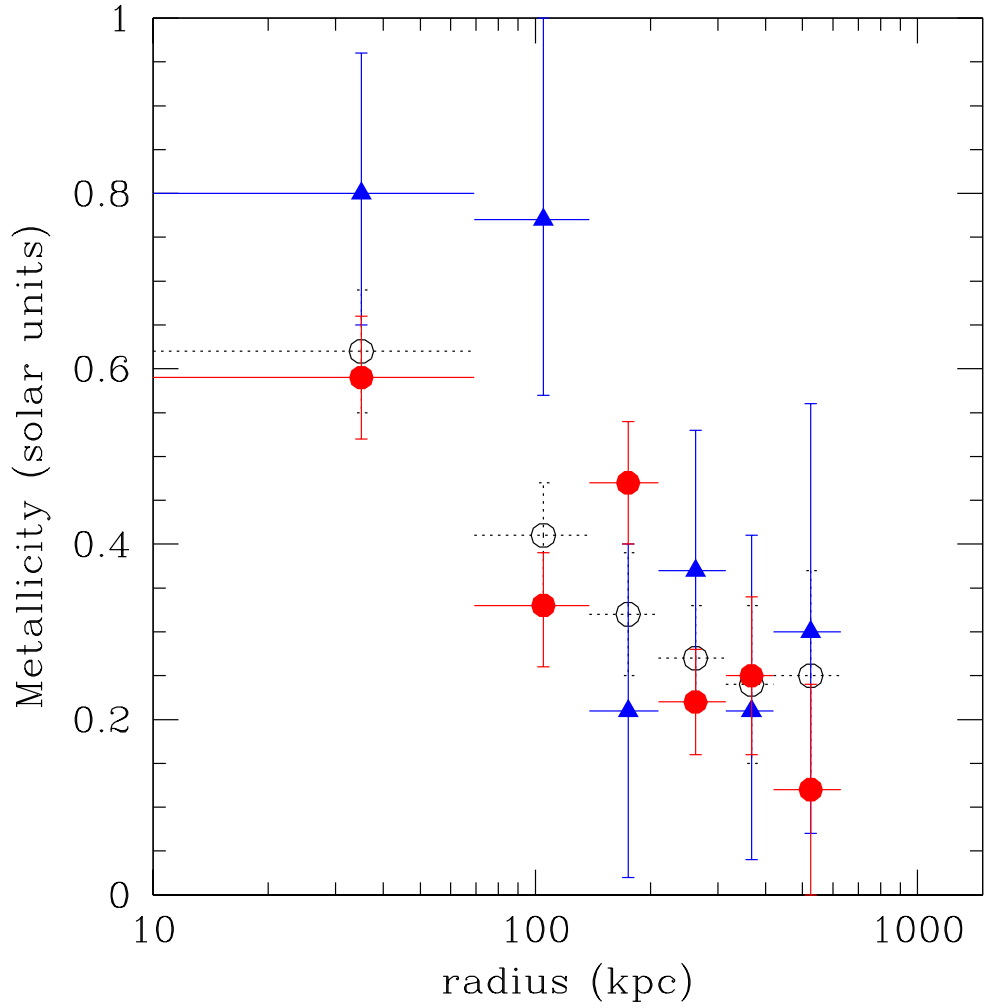}
\caption{ Deprojected X-ray gas metallicity profile measured from MOS1+MOS2 
(blue triangles) and pn (red full circles) spectra. For comparison, 
the projected profile is also shown (black open circles).
The last two bins have been excluded due to poor photon statistics.
\label{profilomet-deproj.fig}}
\end{figure}

The deprojection analysis is performed separately for the
MOS1+MOS2 and pn spectra. The results are reported in Table 2
and the corresponding deprojected
temperature and metallicity profiles are shown in Figs.
\ref{profilot-deproj.fig} and \ref{profilomet-deproj.fig},
respectively.
We also perform the deprojection analysis by
fitting simultaneously the spectra of the three cameras. In
general, the results obtained from pn spectra appear more
reliable, as the deprojected MOS and combined MOS+pn temperature
profiles show some signs of instability and we do not manage to
derive a temperature estimate in the last annulus. In the
following general discussion and in the estimate of the cluster
mass derived from the density and temperature profiles (see Sect.
\ref{mass.sec}) we therefore adopt the pn deprojection results.
As expected, the deprojected central temperature is lower than the
projected one, since in the projected fits the spectrum of the
central annulus is contaminated by hotter emission along the line
of sight.

\begin{figure}
\epsscale{.80}
\plotone{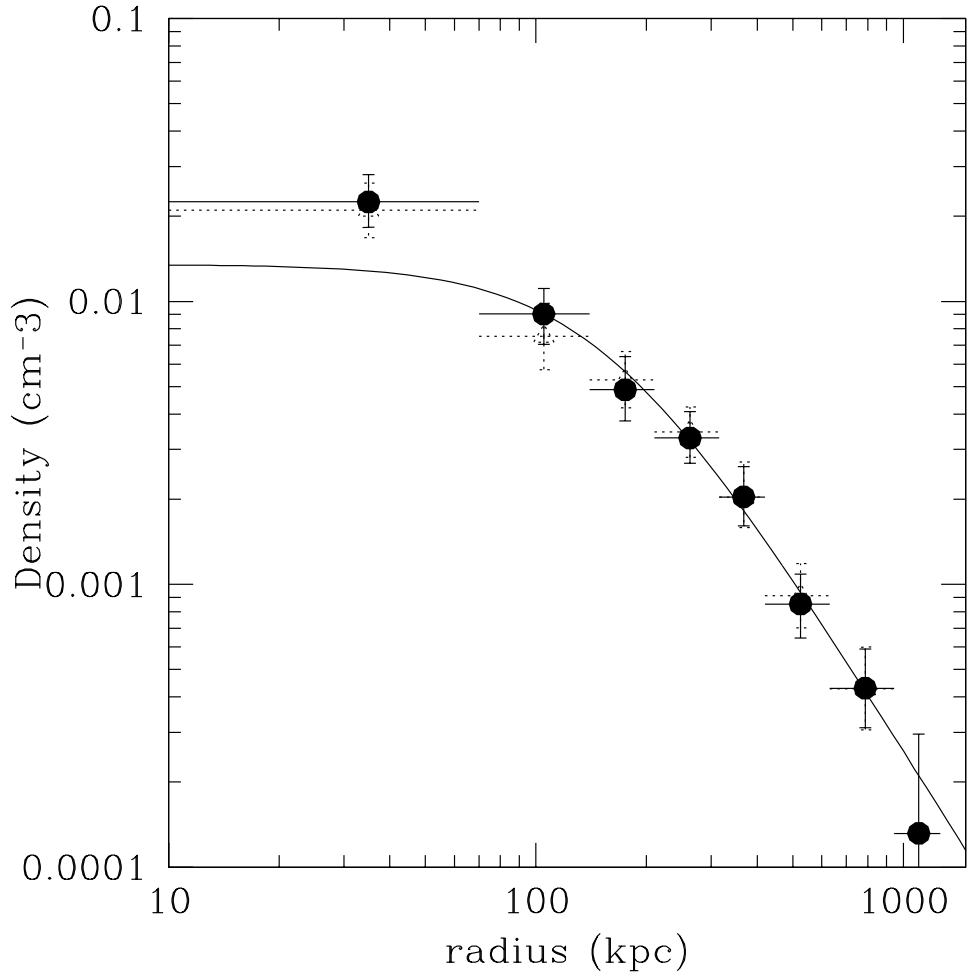}
\caption{Electron density profile measured from pn spectra (full circles). 
For comparison, the profile measured from MOS1+MOS2 spectra is also
shown (open triangles). The solid line indicates the density
profile obtained from the $\beta$-parameters derived by fitting
the surface brightness profile over the 50 - 1000 kpc. See text
for details. 
\label{density.fig}}
\end{figure}

\begin{figure}
\epsscale{.80}
\plotone{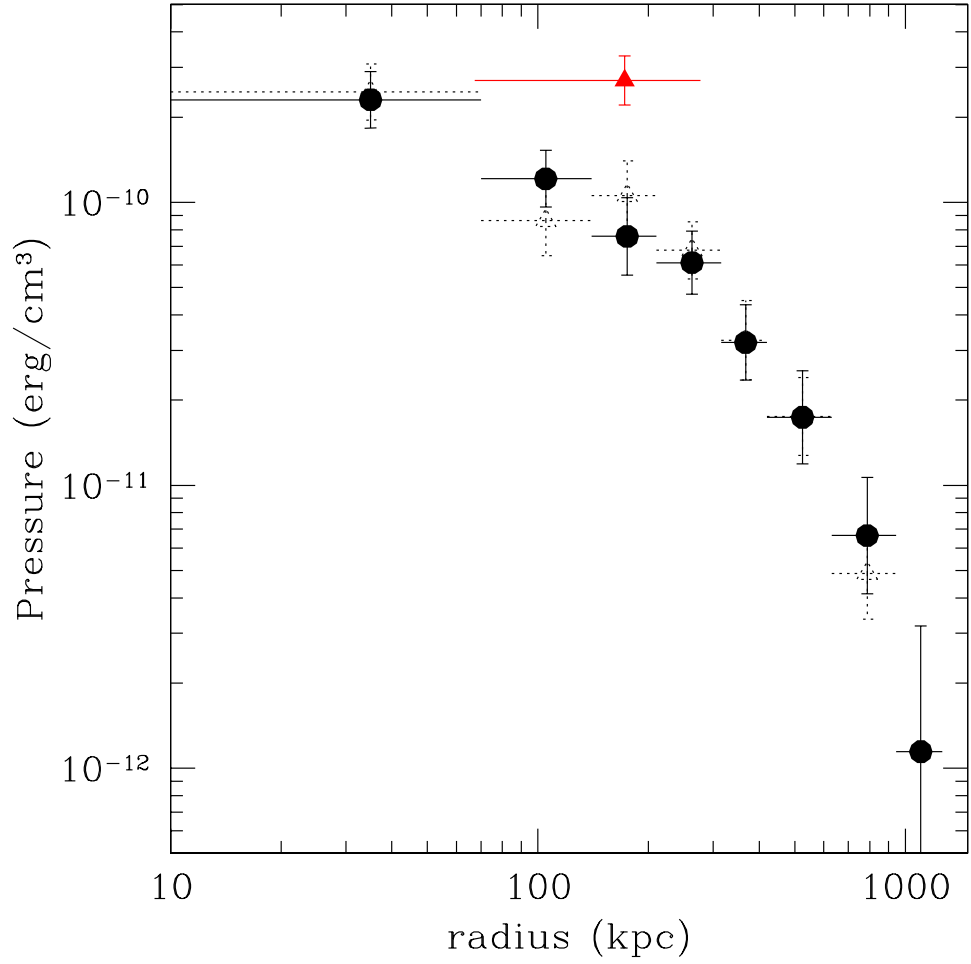} 
\caption{X-ray gas pressure profile measured from pn spectra (full circles). 
For comparison, the profile measured from MOS1+MOS2 spectra is also
shown (open triangles). The red triangle represents the pressure
of the thermal plasma filling the northern cavity, as estimated in
Sect. \ref{cavity.sec}. 
\label{pressure.fig}}
\end{figure}

\begin{figure}
\epsscale{.80}
\plotone{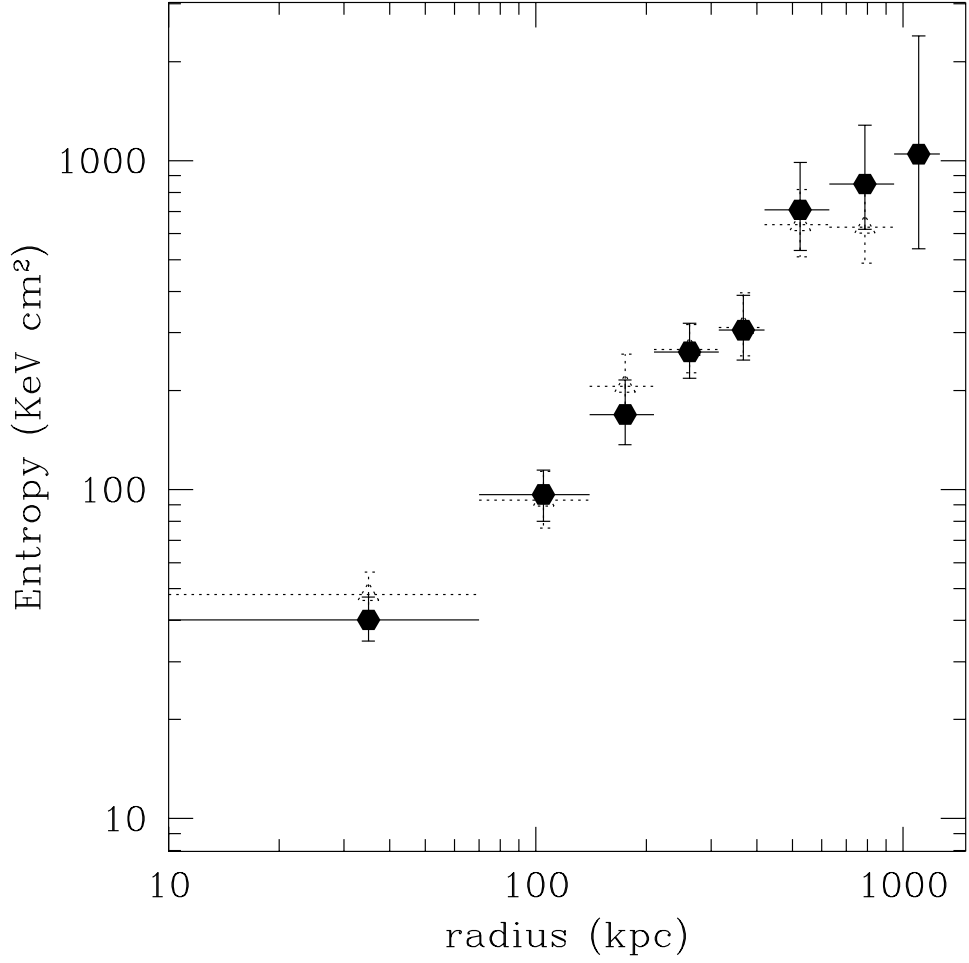}
\caption{X-ray gas entropy profile measured from pn spectra (full circles). 
For comparison, the profile measured from MOS1+MOS2 spectra is also
shown (open triangles). 
\label{entropy.fig}}
\end{figure}

In Figs. \ref{density.fig}-\ref{entropy.fig} we show various
quantities derived from the deprojected spectral fits. The
electron density $n_{\rm e}$ (Fig. \ref{density.fig}) is obtained
from the estimate of the Emission Integral $EI = \int n_{\rm e}
n_{\rm p} dV$ given by the {\ttfamily mekal} normalization:
$10^{-14} EI / ( 4 \pi [D_{\rm A} (1+z)]^2 )$. We assume $n_{\rm
p} = 0.82 n_{\rm e}$ in the ionized intra-cluster plasma. By
starting from the density and temperature information derived from
the deprojection analysis, we can calculate the pressure profile
(Fig. \ref{pressure.fig}) as $P=nkT$, where we assume $n = 2
n_{\rm e}$. The average pressure surrounding the cavities is $\sim
6 \times 10^{-11}$ erg cm$^{-3}$.
Similarly, the entropy profile (Fig. \ref{entropy.fig}) is
calculated from the temperature and density profiles by using the
commonly adopted definition $ S=kT \, n_{\rm e}^{-2/3}$.
Allowing for the different resolutions of \textit{Chandra}
and \textit{XMM}, the high central
entropy level of about 40 keV cm$^2$ that we measure is in
agreement with the value of about 30 keV cm$^2$ measured with
\textit{Chandra} (McNamara et al. 2005).
As pointed out by Voit et al. (2005), this is consistent with being the
result of the accumulation of the very energetic kinetic power outburst at
the cluster center.

We investigate the effect of changing the radial binning by
performing a similar spectral analysis (projected and deprojected)
on different annular regions. We find results consistent with
those presented in Sections \ref{radial.sec} and \ref{deproj.sec}.
We also note that the results presented here are in agreement with those
derived from the analysis of \textit{Chandra} data (McNamara et al. 2005).

\subsection{Cooling Flow}
\label{cf.sec}

MS0735 was identified as a candidate ``cooling flow'' cluster
because of extended $H\alpha$ emission from its central galaxy
(Donahue, Stocke, \& Gioia 1992) and  \textit{ROSAT} HRI detection
of a central high surface brightness peak (Donahue \& Stocke 1995).
However, this identification still has to be confirmed by a detailed
X-ray spectral analysis.

The surface brightness profile, temperature map and temperature
profile derived from our \textit{XMM} data analysis
give indications of the presence of a cooling core. Here
we further investigate the physical properties of the ICM in the
central region. The cooling time is calculated as the time taken
for the gas to radiate its enthalpy per unit volume $H$
using the instantaneous cooling rate at any temperature:
\begin{equation}
t_{\rm cool} \approx \frac{H}{n_{\rm e} n_{\rm H}
\Lambda(T)} = \frac{\gamma}{\gamma -1} \frac{kT}{\mu X_{\rm H} n_{\rm
e} \Lambda(T)} \label{tcool.eq}
\end{equation}
where: $\gamma=5/3$ is the adiabatic index; $\mu \approx 0.61$
(for a fully-ionized plasma) is the molecular weight; $X_{\rm H}
\approx 0.71$ is the hydrogen mass fraction; and $\Lambda(T)$ is
the cooling function. 
We calculate the electron density by following the procedure described in 
Sect. \ref{gas.sec}, using the $\beta$-parameters derived by fitting
the surface brightness profile over the 50 - 1000 kpc region\footnote{
The best fit obtained in Sect. \ref{brightness.sec} cannot be
extrapolated to the central region ($r$ \ltsim 100 kpc) therefore cannot 
be used here for the purpose of calculating the central cooling time.}
(data in this region can be approximated by a $\beta$ model with $r_{\rm
c} \sim 150$ kpc and a slope parameter $\beta \sim 0.70$). 
We note that in the radial range where the $\beta$ model is a good
representation of the observed surface brightness profile, the
density profile derived from this method agrees with that derived
from the deprojection analysis of spectra extracted in concentric
annuli (see Fig. \ref{density.fig}).

Following B\^{\i}rzan et al. (2004), we define the cooling radius as
the radius within which the gas has a cooling time less than $7.7
\times 10^9$ yr, the look-back time to $z=1$ for our adopted cosmology. 
With this definition, we find $r_{\rm cool} \sim 100$ kpc which corresponds 
to the central 30 arcsec. The accumulated  spectrum within this radius
is extracted and compared to three  different spectral models.
Model A is the {\ttfamily mekal} model already used in Sect. \ref{global.sec}.
Model B includes a single temperature component plus an isobaric
multi-phase component ({\ttfamily mekal + mkcflow} in XSPEC),
where the minimum temperature, $kT_{\rm low}$, and the
normalization of the multi-phase component, Norm$_{\rm low} =
\dot{M}$, are additional free parameters. This model differs from
the standard cooling flow model as the minimum temperature is not
set to zero. Finally, in model C the constant pressure cooling
flow is replaced by a second isothermal emission component
({\ttfamily mekal + mekal} in XSPEC). As for model B, this model
has 2 additional free parameters with respect to model A: the
temperature, $kT_{\rm low}$, and the normalization, Norm$_{\rm
low}$, of the second component.

The results, summarized in Table \ref{cf.tab}, show that the
statistical improvements obtained by introducing an additional
emission component (models B or C) compared to the
single-temperature model (model A) are significant at more than
the 99\% level according to the F-test. With our data, however, we
cannot distinguish between the two multi-phase models. This means
that the extra emission component can be equally well modelled
either as a cooling flow or a second isothermal emission
component. We note that the fit with the modified cooling flow
model sets tight constraints on the existence of a ``pedestal'' minimum
temperature ($\sim$ 1.5 keV). The nominal mass deposition rate in
this empirical model is $\sim 260 \pm 30 \mbox{ M}_{\sun} \mbox{
yr}^{-1}$. 
We also attempt the fit with a classical cooling flow model 
(model D), by imposing $kT_{\rm low}= 0.1$ keV, and find 
a mass deposition rate ${\dot M} \sim 40 \pm 10
\mbox{ M}_{\sun} \mbox{yr}^{-1}$.


\subsection{Cavity regions}
\label{cavity.sec}

Several states of matter have been proposed to fill the cavities
(Pfrommer et al. 2005), including a population of relativistic
electrons radiating at low radio frequencies and a dilute,
shock-heated thermal gas.
The synchrotron emission from radio-filled cavities provides evidence
of the existence of relativistic electrons and magnetic fields.
The cavities may also be filled with a shock-heated thermal gas that
can contribute to their internal pressure.
This possibility is strengthened if the assumption of equipartition holds.
Indeed, since the equipartition estimates of the nonthermal pressure
in the radio bubbles give values which are typically a factor of ten
smaller than the thermal pressures of the surrounding X-ray gas   
(e.g., Blanton et al. 2001, De Young 2006), 
the fact that the cavities are long-lived indicates that the necessary 
pressure support might be supplied by an additional thermal component.
Observationally, besides the detection of hot X-ray emitting gas 
claimed by Mazzotta et al. (2002) within the ghost cavity of
the MKW3s cluster, there has been no detection of hot gas over the cavities
even in the clusters with better data available.

\begin{figure}
\epsscale{.80}
\plotone{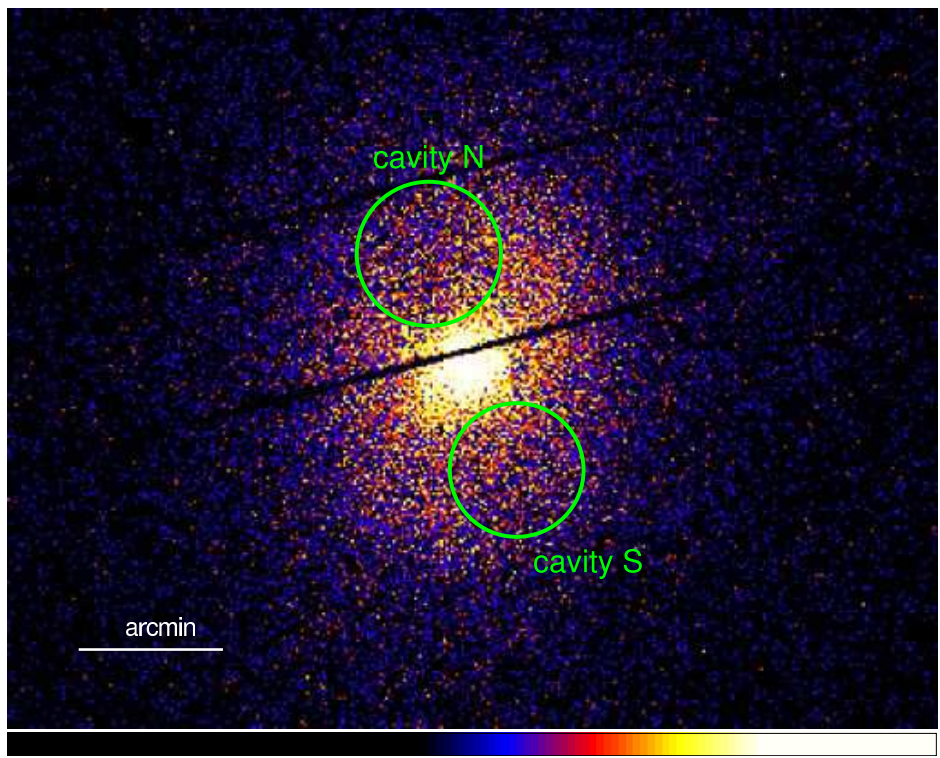}
\caption{
Regions considered for the spectral analysis of the cavities.
The N and S cavities have a radius of  105 kpc and
are located at a distance of $\sim 170$ and $\sim 180$
kpc from the cluster center, respectively.
\label{cavities.fig}}
\end{figure}

In the case of MS0735, the existence of
a relativistic plasma filling the cavities is clearly indicated by
the presence of the radio source at the position coincident with
the holes in the X-ray emission (McNamara et al. 2005).
In order to investigate the possibility of an additional thermal gas
component in the cavities,
we perform a detailed analysis by modeling the spectra extracted in
the cavity regions as the sum of ambient cluster emission and a
hot thermal plasma, each with a characteristic temperature. The
northern and southern cavity regions, detailed in Fig.
\ref{cavities.fig}, contain $10083$ and $8872$ source
counts with a count rate of 0.23 and 0.20 counts/s (MOS+pn),
respectively. The results of the spectral analysis are reported in
Table \ref{cavity.tab}.

In the northern cavity, which shows a higher brightness contrast
than the southern one in the \textit{XMM} image, we find
indications of the presence of a $\sim 13$ keV plasma. By assuming
that the emission of this hottest component comes entirely from
the cavity region whereas that of the coolest component is due to
the projected foreground and background cluster emission\footnote{
This assumption is justified by the fact that a
component as hot as 13 keV is detected nowhere else in the cluster
and therefore is plausibly located only in the cavity region.}, 
we can estimate the density of the thermal plasma in the cavity to be
$n_{\rm e} \sim 6.4 \times 10^{-3}$ cm$^{-3}$. This leads to an
estimate of its pressure of $\sim 2.7 \times 10^{-10}$ erg
cm$^{-3}$, which is a factor $\sim$ 3-4 higher than the pressure
of the surrounding medium (see Fig. \ref{pressure.fig}).
However, we note that the 
improvement in the fit due to the extra thermal component is only 
marginally significant\footnote{$\chi^2$/dof = 385/311 compared to 
$\chi^2$/dof = 394/313, with F-test probability=0.028.}. 
We therefore cannot place strong constraints on the existence of a hot 
thermal component filling the cavities and the estimate presented above
has to be considered with caution.

\subsection{Shock front}
\label{shock.sec}

\begin{figure}
\epsscale{.80}
\plotone{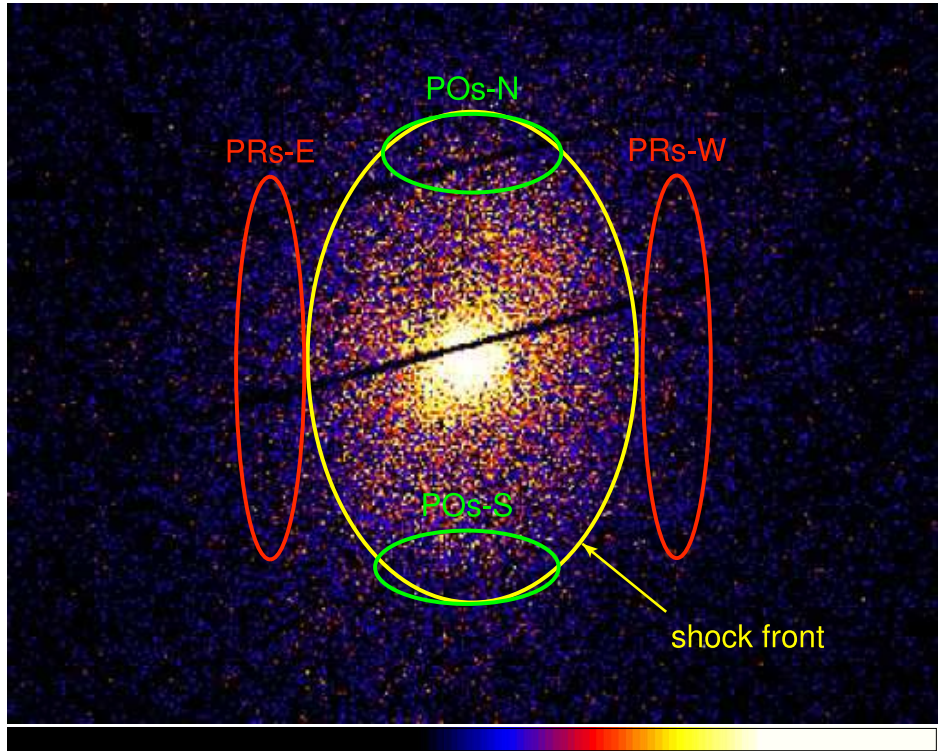}
\caption{Elliptical regions considered for the spectral analysis of the
pre-shock and post-shock gas. 
The centers of the ellipses are chosen to be at
the same radial distance from the cluster center ($\sim 300$ kpc),
in order to allow us to make a direct comparison of the spectral
properties in different directions with respect to the azimuthally-averaged 
profiles (see Figs. \ref{temp-shock.fig} and \ref{met-region.fig}).
\label{shock-a.fig}}
\end{figure}

\begin{figure}
\epsscale{.80}
\plotone{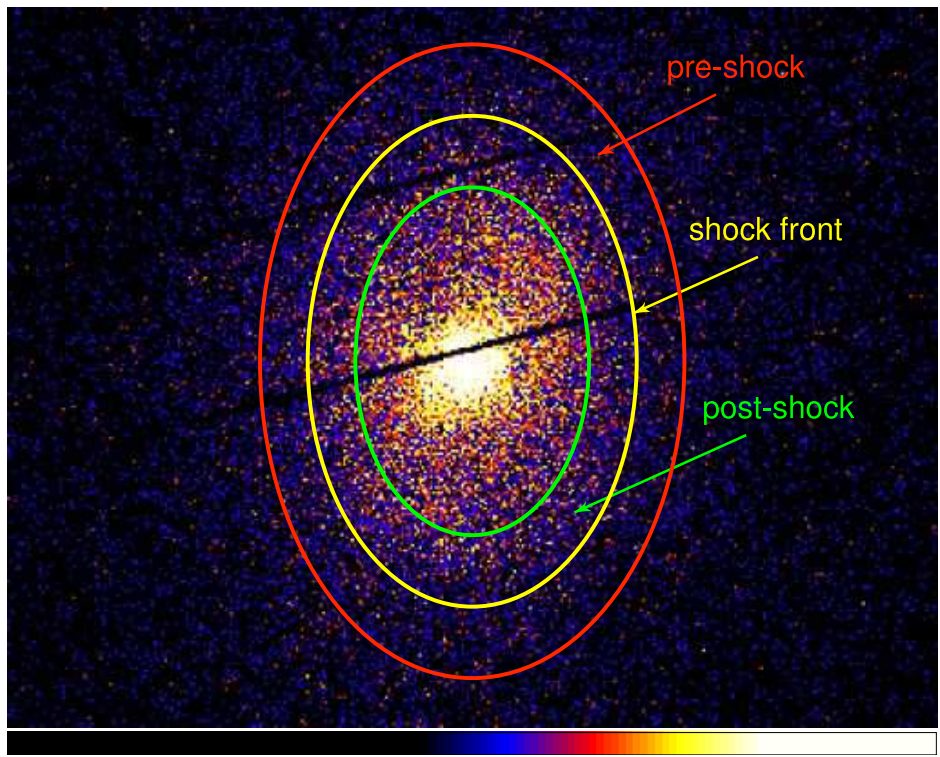}
\caption{Elliptical regions considered for the spectral analysis of the
pre-shock and post-shock gas. 
The concentric annular regions are defined in order to surround the whole 
shock front as derived from the \textit{Chandra} data.
\label{shock-b.fig}}
\end{figure}

\textit{Chandra} observation of MS0735 reveals a discontinuity in the X-ray
surface brightness that has been interpreted as a weak cocoon
shock driven by the expansion of the radio lobes that inflate the
cavities. The shock front has an elliptical shape, being located
$\sim 240$ kpc and $\sim 360$ kpc from the cluster center in
east-west and north-south directions, respectively (McNamara et
al. 2005).
We find indications of a surface brightness feature at
a position consistent with that seen in the \textit{Chandra} data,
however we cannot clearly distinguish a break in surface brightness that
would be a signature of the shock front.
By considering the elliptical shock front position derived from
\textit{Chandra} data, we perform a spectral analysis on the regions
detailed in Figs. \ref{shock-a.fig} and \ref{shock-b.fig} in order to
derive the temperature of the pre-shock and post-shock gas. The
results are reported in Table \ref{shock.tab}.

\begin{figure}
\epsscale{.80}
\plotone{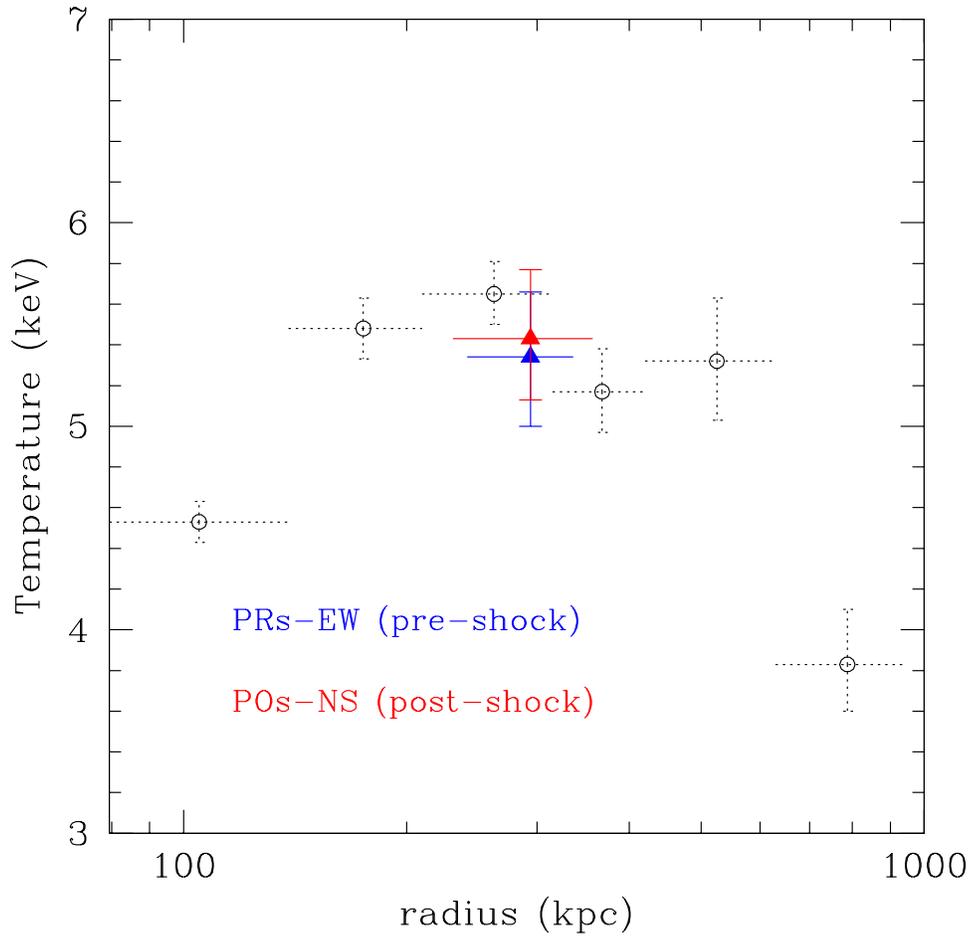}
\caption{The open circles represent the projected temperature profile
measured in Sect. \ref{radial.sec}. 
The blue triangles indicate the measurements in the pre-shock regions 
along the east-west direction,  
the red square indicate the measurements in the post-shock regions 
along the north-south direction (see Fig. \ref{shock-a.fig}).
The error bars are at 1$\sigma$ level.
\label{temp-shock.fig}}
\end{figure}

\begin{figure}
\includegraphics[angle=-90,scale=.70]{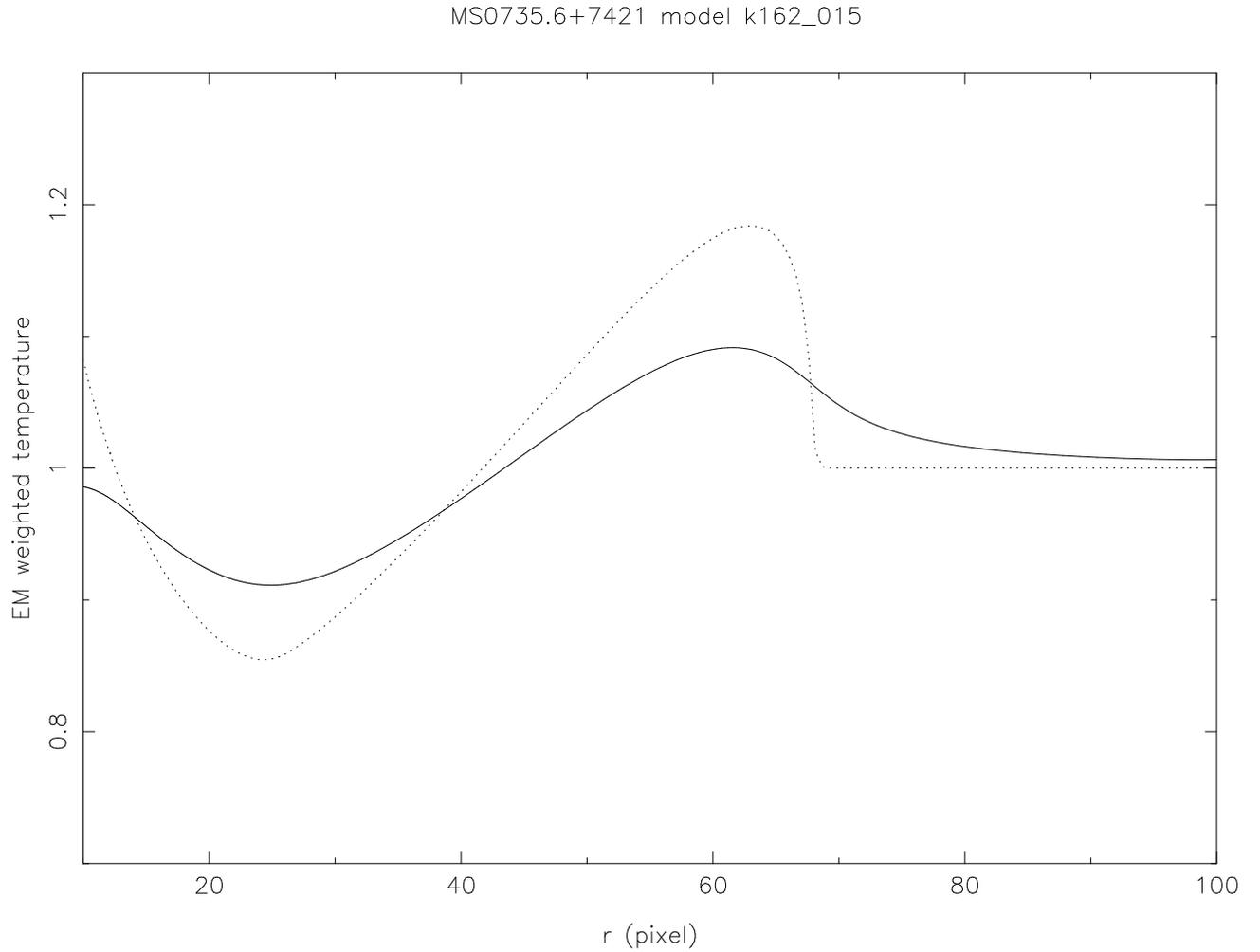}
\caption{
Temperature jump expected by the shock model. The dotted line shows the
projected temperature profile as it would be seen by \textit{Chandra}, and
the full line is the result after smearing with the \textit{XMM}
PSF.  The temperatures are in units of the pre-shock temperature,
assumed to be constant. The radial distance is in \textit{Chandra} pixel
unit (1 pixel = 0.492 arcsec).
\label{shock-model-a.fig}}
\end{figure}

\begin{figure}
\epsscale{.80}
\plotone{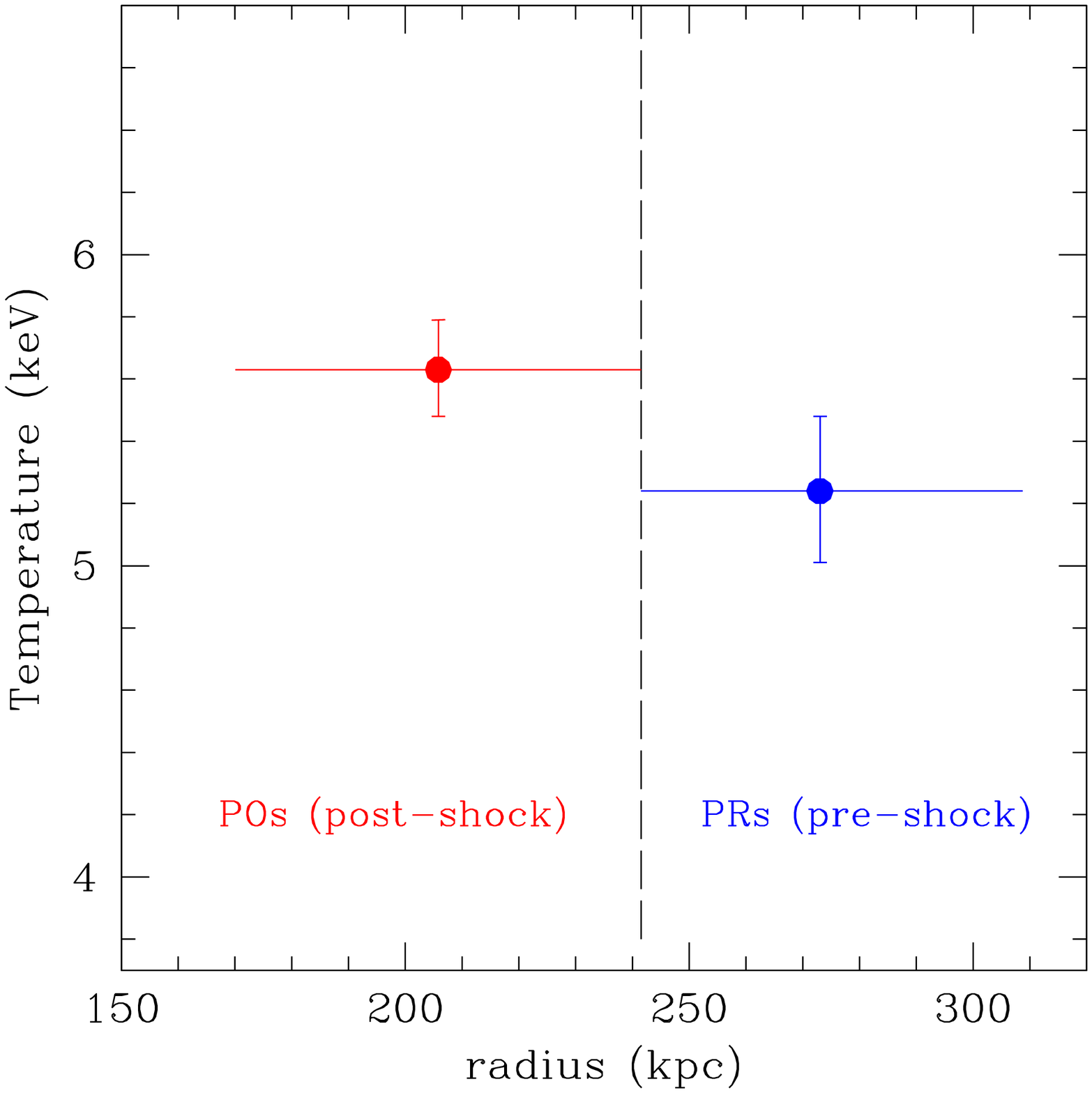}
\caption{
Temperature measured in the annular pre-shock (blue) and
post-shock (red) regions (see Fig. \ref{shock-b.fig}).
The dashed line represents the position
of the shock front along the east-west direction. The error bars
are at 1$\sigma$ level.
\label{shock-model-b.fig}}
\end{figure}

The particular choice of the elliptical regions next to specific
sectors of the shock front (see Fig. \ref{shock-a.fig})
allows us to compare directly the spectral properties of the
pre-shock and post-shock regions in different directions with
respect to the azimuthally-averaged profiles. 
We do not
find any appreciable variation in temperature (see Fig.
\ref{temp-shock.fig}).
In the analysis of the \textit{Chandra} data, the shock properties are
determined using a spherical hydrodynamic model of a point
explosion at the center of an initially isothermal, hydrostatic
atmosphere. When adapted to the \textit{XMM} data by taking into
account the smearing effect due to the large PSF, such a model
predicts a temperature rise of $\sim$ 10\% (see Fig.
\ref{shock-model-a.fig}). Our spectral results obtained from
elliptical annular regions next to the whole shock front (see
Fig. \ref{shock-b.fig}) do suggest a temperature jump at the
shock front, in agreement with those derived from previous \textit{Chandra}
observations, although due to the large error bars the pre-shock
and post-shock regions are still consistent with being isothermal
(see Fig. \ref{shock-model-b.fig}). 
From this analysis we therefore conclude that the \textit{XMM} data can
neither confirm or deny the existence of the shock front. 

\section{Mass profile}
\label{mass.sec}

\subsection{Total gravitational mass}

In the following we estimate the total mass of the cluster using the
usual assumptions of hydrostatic equilibrium and spherical symmetry.
Under these assumptions, the gravitational mass $M_{\rm tot}$ of a
galaxy cluster can be written as:
\begin{equation}
M_{\rm tot}(<r) = - \frac{kT \, r}{G \mu m_{\rm p}}
\left[ \frac{ d \ln n_g}{d \ln r} + \frac{d \ln T}{d \ln r} \right]
\label{mass.eq}
\end{equation}
where $G$ and $m_{\rm p}$ are the gravitational constant and proton mass and
$\mu \approx 0.61$.
The mass contribution from galaxies is small, thus we neglect it.
Therefore
$
M_{\rm tot} (<r) = M_{\rm gas} (<r) + M_{\rm DM} (<r)
$,
i.e. the total gravitational mass within a sphere of radius $r$ is given
by gas plus dark matter mass.

The mass profile derived from the equilibrium equation is strongly
dependent upon the measured temperature profile. Large errors and irregular
radial distribution of the temperature values induce large scatter on the
reconstructed gravitational mass measurements.
In order to obtain reliable mass estimates it is therefore crucial to
select bins with a robust temperature estimate.
For this reason we exclude the last temperature bin, although its
inclusion would allow us to trace the temperature profile and therefore
mass profile at much larger radii.

\begin{figure}
\epsscale{.80}
\plotone{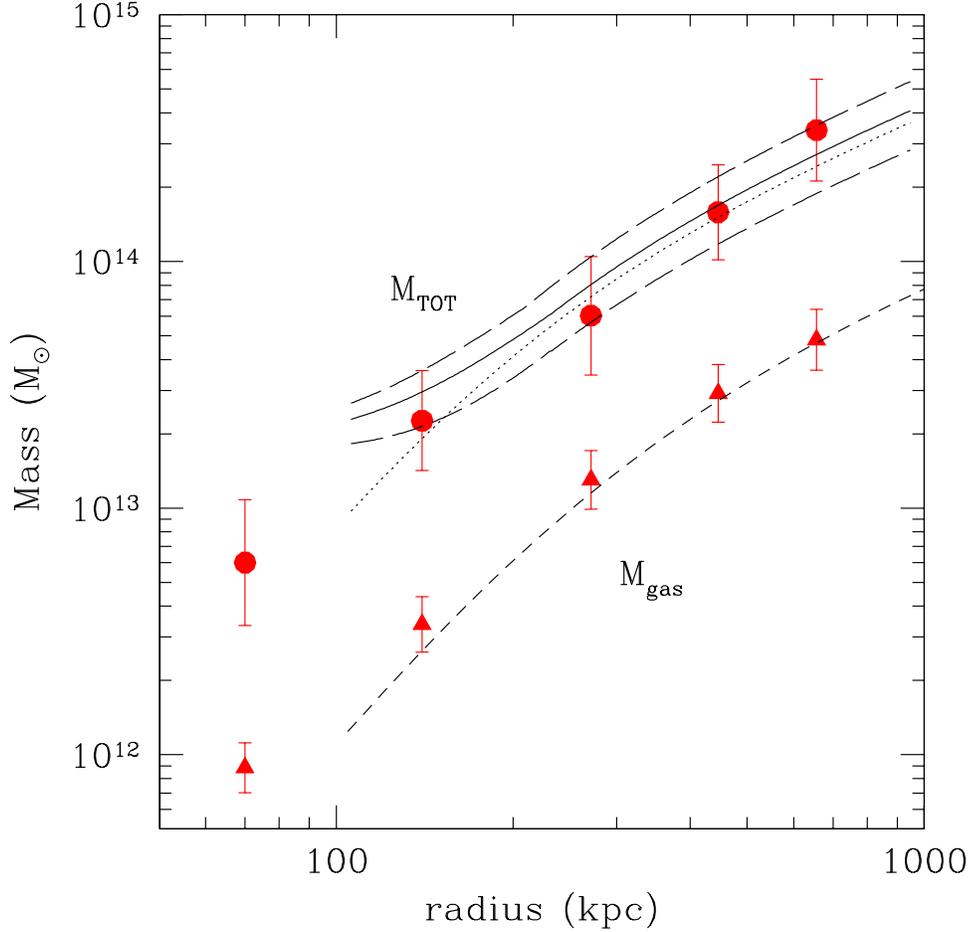}
\caption{Mass profiles calculated with different methods.
\textit{Red circles:} total mass profile derived from deprojected gas
temperature and density values.
\textit{Black solid line:} total mass calculated from
Eq. \ref{massbeta.eq} by estimating the temperature gradient from the
deprojected temperature profile
(with error on the mass calculation coming from the
temperature measurement and $\beta$-model shown by the \textit{black long
dashed lines}).
\textit{Black dotted line:} total mass calculated from
Eq. \ref{massbeta.eq} by assuming a constant temperature of 4.79 keV.
\textit{Red triangles:} gas mass profile derived from deprojected analysis.
\textit{Black dashed line }: gas mass profile derived from $\beta$-model fit.
See text for details.
\label{mass.fig}}
\end{figure}

The deprojected $d \ln n_g/d \ln r$ is calculated from
the parameters of the $\beta$-model derived in Sect. \ref{brightness.sec}.
In particular,
the advantage of using a $\beta$-model to parameterize the observed surface
brightness is that gas density and total mass profiles can be recovered
analytically and expressed by simple formulae:
\begin{equation}
n_{\rm gas}(r) = n_{\rm 0,gas} \left[ 1+ \left(\frac{r}{r_{\rm c}} \right)^2
\right]^{-3 \beta/2}
\label{n.eq}
\end{equation}
\begin{equation}
M_{tot}(<r) = \frac{k \, r^2}{G \mu m_p} \left[
\frac{ 3 \beta r T}{r^2 + r^2_{\rm c}} - \frac{d T}{d r} \right]
\label{massbeta.eq}
\end{equation}
As a first-order approximation, the temperature gradient is estimated by
dividing the deprojected temperature profile (see Fig.
\ref{profilot-deproj.fig}) into two radial intervals and least-squares 
fitting a straight line in each interval.
The total gravitating mass distribution derived from Eq.
\ref{massbeta.eq} is shown in Fig. \ref{mass.fig} as a solid line,
with errors coming from uncertainties in the temperature measurement and
$\beta$-model parameters represented as long dashed lines.
In Fig. \ref{mass.fig} we also show (dotted line) the total mass calculated
by assuming a constant temperature of
$<T_{\rm X}> = 4.79$ keV, where the average emission-weighted cluster 
temperature $<T_{\rm X}>$ is derived by extracting the global spectrum 
of the cluster after excluding the cooling flow region (i.e. in the annular 
region 30 arcsec - 6 arcmin).
We note that the total integrated mass within a particular volume is
dependent upon the local physical properties 
(local temperature and density gradients) 
and is is not strongly affected by the regions interior, or
exterior, to that radius.
The mass profile derived with this method is thus reliable 
in the region where the $\beta$-model is a good representation of the
observed surface brightness profile (100 kpc \ltsim \, r \, \ltsim 1150 kpc,
see Sect. \ref{brightness.sec}), whereas it cannot be extrapolated to the
central region.

As an alternative method,
we then calculate the total mass by making direct use of the deprojected
gas temperature and
electron density values estimated from the spectral best-fit with a single
phase model (Sect. \ref{deproj.sec}).
In particular,
the total mass enclosed within the midpoint between two consecutive shells
is calculated from the pressure gradient and the gas density at the midpoint
by following the method described in Voigt \& Fabian (2006).
The mass profile reconstructed by using this method, is shown as full red
circles in Fig. \ref{mass.fig}.
The uncertainties are calculated using error propagation from the pressure
and density estimates (90\% confidence level).
We note that the cavities may affect the mass determination.
In particular, since the cluster does not exhibit a smooth pressure
profile between 100-400 kpc (see Fig. \ref{pressure.fig}), which would
result in a non-monotonically increasing mass profile in this region, we
exclude the fourth data point in the pressure and gas density profiles.

The mass measurements reconstructed through the two
different methods (direct application of the hydrostatic equilibrium equation
with density profile derived from either $\beta$-model fit to the surface
brightness profile or deprojection analysis on annular spectra) are
in very good agreement (see also Fig. \ref{nfw.fig}).

\subsection{Gas mass and gas mass fraction}
\label{gas.sec}

We estimate the gas mass
through two different methods, i.e. by using the density profile
derived from either the $\beta$-model fit to the surface
brightness profile or the deprojection analysis on annular spectra.
To convert from electron number density to gas density we use the relation
$\rho = 1.83 \mu m_{\rm H} n_{\rm e}$.

In the first case, the gas mass profile is
derived by integrating the gas density given by Eq. \ref{n.eq} in spherical
shells and using the $\beta$-model parameters determined in Sect.
\ref{brightness.sec}.
The normalization of Eq. \ref{n.eq} is obtained from the combination
of the best-fit results from the spectral and imaging analyses, which
allows us to determine the conversion count rate - flux used to derive the
bremsstrahlung emissivity that is then integrated along the line-of-sight
and compared with the central surface brightness value.
The adoption of the parameters of the $\beta$-model fit in
the outer regions produces an underestimate of the derived central electron
density (see Fig. \ref{density.fig}) and this turns
into an underestimate of the gas mass in the central shells.
Since the integrated gas mass is calculated by summing from the center
outwards, any error in the measurement at small radii will propagate out
to larger radii. However, the gas mass at small radii is much less than at
large radii and any uncertainty in the measurements in the core are unlikely
to have a significant effect on the gas mass profile further out.
The gas mass profile derived from the $\beta$-model is shown as a dashed line
in Fig. \ref{mass.fig}.
The gas mass profile obtained by integrating in spherical shells the gas
density derived directly from the deprojected electron number density
is shown in Fig. \ref{mass.fig} as red triangles.
Again, we note that the two profiles are in very good agreement.

\begin{figure}
\epsscale{.80}
\plotone{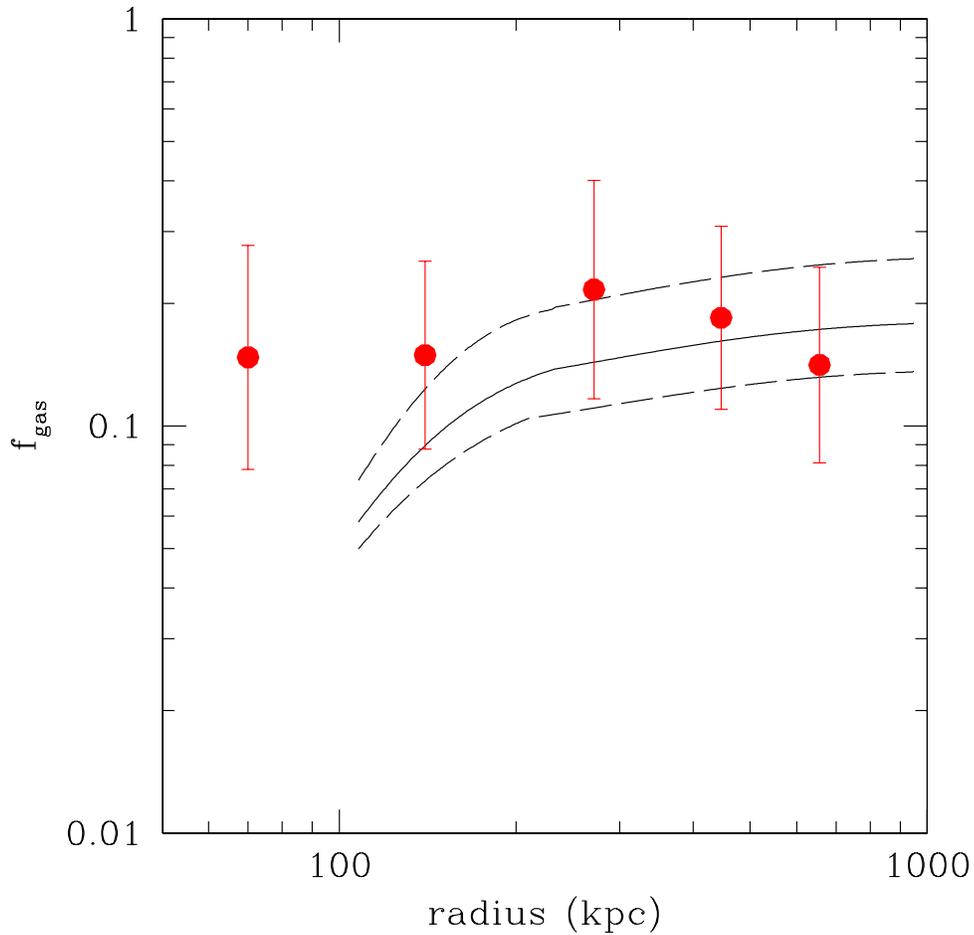}
\caption{
$f_{\rm gas}$ profile derived with different methods.
\textit{Red circles:} $f_{\rm gas}$ profile derived from $M_{\rm tot}$ and
$M_{\rm gas}$ measured directly from the deprojected density and temperature
values (see red circles and red triangles in Fig. \ref{mass.fig}).
\textit{Black solid line:} $f_{\rm gas}$ profile (with error shown by the
\textit{black long dashed lines}) derived from $M_{\rm tot}$ and
$M_{\rm gas}$ measured by using the density profile derived from
the $\beta$-model fit to the surface brightness profile
(see black solid line and black dashed line in Fig. \ref{mass.fig}).
\label{fgas-mes.fig}}
\end{figure}

The gas mass fraction is the ratio of the total gas mass
to the total gravitating mass within a fixed volume:
\begin{equation}
f_{\rm gas}(r) = \frac{M_{\rm gas}(< r)}{M_{\rm tot}(< r)}
\label{fgas}
\end{equation}
The gas mass fraction profile derived from $M_{\rm tot}$ and $M_{\rm gas}$
measured from the $\beta$-model and deprojection methods
are plotted in Fig. \ref{fgas-mes.fig} as black lines and red points, respectively.
Adding the mass contribution from galaxies 
($\sim$1-2\%, Lin et al. 2003)
to the total mass would have a
small effect on the estimate of $f_{\rm gas}$, with variations lying within
the error bars.

\section{Discussion}
\label{discussion.sec}

Simple assumptions for the formation and evolution of galaxy clusters
predict self-similar scaling relations.
Galaxy clusters are assumed to form by spherical collapse of
dark matter and gas due to gravitational instability. 
The gas is heated by shocks and adiabatic compression.
For dissipationless collapse, the X-ray emitting gas can be
considered to share the same properties as the dark matter
(e.g. Bryan \& Norman 1998; Arnaud \& Evrard 1999).
Self-similar dark matter haloes are assumed to have identical properties when
scaled by the virial radius, which is the radius
separating the region where the cluster is in hydrostatic equilibrium
from where matter is still infalling.
However, deviations from self-similarity are expected in the ICM due
to the details of individual collapse histories and to any physical
processes beyond simple gravity and gas dynamics (e.g. Evrard \& Henry
1991; Bryan \& Norman 1998; Borgani et al. 2002 and references therein).
In particular, the ICM of cooling flow clusters is affected significantly
by heating that boosts the thermal energy of the gas.

Here we investigate the correlations between the physical quantities
observed for MS0735 in order to scale it in the overall populations of clusters
and assess its conformity to the generally observed mass-temperature
($M$-$T$) and luminosity-temperature ($L$-$T$) relations.
Our final aim is to evaluate the impact of energetic AGN outbursts
on such scaling relations, which are the foundation to construct
the cluster mass function and to use these virialized objects as cosmological
probes.


\subsection{Modeling the observed mass profile: determination of
$r_{\Delta}$}
\label{rdelta.sec}

In order to rank MS0735 in the overall population of clusters we investigate
the relations among different physical quantities considering their values
at a given overdensity $\Delta$.
This is defined with respect to the critical density
$
\rho_{{\rm c},z} = (3 H_z^2)/ (8 \pi G)
$,
and within a cluster described as a sphere with radius $r_{\Delta}$:
\begin{equation}
\Delta = \frac{3 M_{\rm tot}(<r_{\Delta})}{4 \pi \rho_{{\rm c},z}
r^3_{\Delta}}
\label{delta.eq}
\end{equation}
where the Hubble constant at redshift z is equal to (e.g., Bryan \& Norman
1998):
\begin{equation}
H_z = H_0 \sqrt{\Omega_m (1+z)^3 + 1 - \Omega_m} \equiv H_0 \; h(z)
\label{hz.eq}
\end{equation}
In order to estimate $r_{\Delta}$ for various overdensities we need to
fit the mass profile with an analytical function, that allows us to
extrapolate the mass profile beyond the outer radius accessible to our
X-ray observations.
As a cosmologically motivated dark matter mass model, we consider the
integrated NFW (Navarro et al. 1996) dark matter profile:
\begin{equation}
M_{\rm DM} (<r) = 4 \pi r^3_{\rm s} \rho_{\rm c,z} \frac{200}{3}
\frac{c^3 \left( \ln (1+r/r_{\rm s}) - \frac{r/r_{\rm s}}{1 + r/r_{\rm s}}
\right)}{ \ln (1+c) - c/(1+c)}
\label{nfw.eq}
\end{equation}
The scale radius $r_{\rm s}$ and the concentration parameter $c$ are the
free parameters.

\begin{figure}
\epsscale{.80}
\plotone{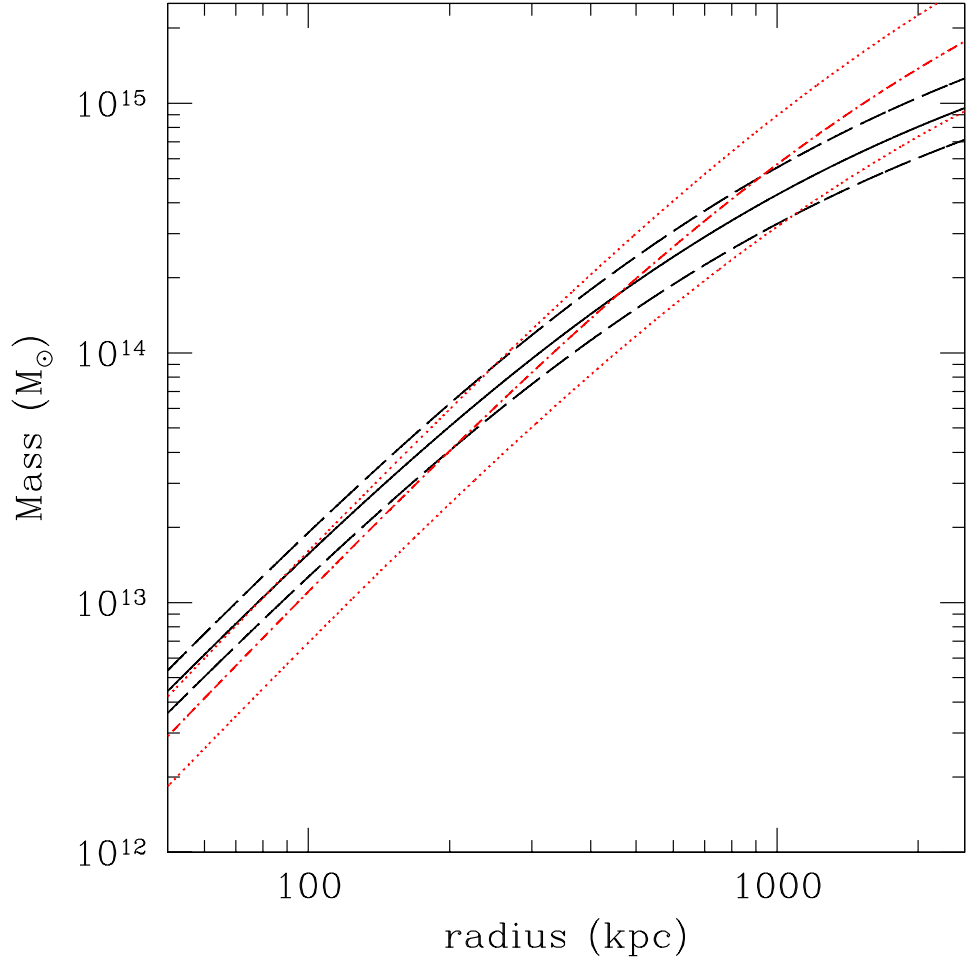}
\caption{
Best fits of the NFW mass profile calculated with different methods.
\textit{Red dashed-dotted line:}
best-fit NFW to the total mass profile measured from the deprojected
density and temperature values,
with $1 \sigma$ error shown by the \textit{red dotted lines}.
\textit{Black solid line:}
best-fit NFW to the total mass profile measured from the density profile
derived from the $\beta$-model fit to the surface brightness profile,
with $1 \sigma$ error shown by the \textit{black long dashed lines}.
\label{nfw.fig}}
\end{figure}

\begin{figure}
\epsscale{.80}
\plotone{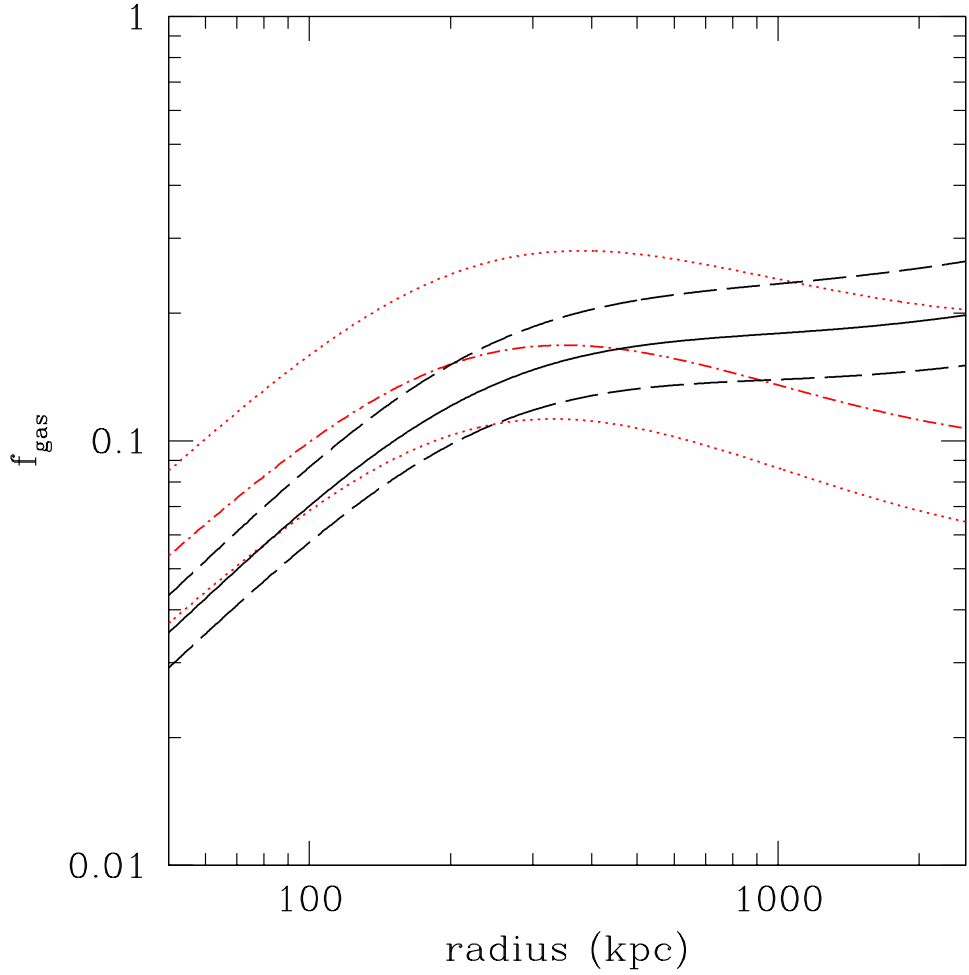}
\caption{
$f_{\rm gas}$ profile estimated from the NFW total mass profile derived with
different methods (the gas mass is estimated from
the $\beta$-model fit to the surface brightness profile).
The line colors and styles are the same as in Fig. \ref{nfw.fig}.
\label{fgas-nfw.fig}}
\end{figure}

We perform the fit to minimize the $\chi^2$ of the comparison
between the mass predicted by Eq. \ref{nfw.eq} and the mass profile
reconstructed from both the density profiles estimated from
the $\beta$-model and deprojection methods.
The best fit parameters for the mass profile
reconstructed from density values estimated from the $\beta$-model
are:
$r_s = 498^{+36}_{-38}$ kpc, $c = 3.45^{+0.18}_{-0.19}$
($\chi^2/$dof = 0.08/4).
The best fit parameters for the mass profile
reconstructed from deprojected density values are:
$r_s = 1343^{+235}_{-240}$ kpc, $c = 1.66^{+0.18}_{-0.21}$
($\chi^2/$dof = 0.18/3).
The quoted error are at the 68\% confidence levels (1$\sigma$).
The comparison of the two best fits of the NFW mass profile is shown
in Fig. \ref{nfw.fig}.
The corresponding gas mass fraction estimated from the NFW mass profile fit
is shown in Fig. \ref{fgas-nfw.fig} for both the $\beta$-model and
deprojection methods.
Note that we neglect the gas mass contribution to the total mass and we
assume
$
M_{\rm tot} (<r) = M_{\rm DM} (<r)
$.
We also perform the same fitting procedure by including the gas mass,
i.e. by assuming
$
M_{\rm tot} (<r) = M_{\rm DM} (<r) + M_{\rm gas} (<r)
$,
and find very similar results.

From the best fit parameters we compute $r_{\Delta}$ from Eq. \ref{delta.eq}
for various overdensities: $\Delta = 2500, 1000, 500, 200$.
For $\Omega_{m} + \Omega_{\Lambda} =1$, the virial radius
corresponds to $r_{\Delta}$ if
$\Delta = \Delta_{\rm vir} = 178 \Omega_{\rm m}^{0.45}$ (Lacey \& Cole
1993; Eke et al. 1996, 1998).
For $\Omega_{\rm m} = 0.3$ adopted here, $\Delta_{\rm vir} = 104$.
The virial radius derived in an Einstein-de Sitter Universe
($\Omega_{\rm m} = 1$) is still often used in the literature,
corresponding to a density contrast of
$\Delta_{\rm vir} = 178$  (or $\Delta_{\rm vir} = 200$), so that
$r_{\rm vir} \sim r_{\rm 200}$.
The relation $r_{\rm vir} = c \times r_s$ holds for the NFW mass profile.
The results obtained from the NFW fit to the mass estimated through the
$\beta$-model and deprojection methods are shown in the Tables
\ref{rdelta-beta.tab} and \ref{rdelta-deproj.tab}, respectively.
We also report the total mass and the gass mas fraction.
The error related to the mass estimate is obtained from half the difference
between the maximum and the minimum value calculated at each radius for the
set of parameters acceptable at $1 \sigma$.
The gas mass fraction is calculated at
given overdensities by adopting the gas mass computed from the integration
of the $\beta$-model fit to the density profile extrapolated out to
$r_{\Delta}$. The reported error is obtained from the error propagation and is
dominated by the error on the total mass.

We note that the value obtained for $r_{\rm 200}$ is highly dependent on the
accuracy of $r_{\rm s}$, since it involves extrapolating the NFW model
using $r_{\rm s}$. The measure of  $r_{\rm s}$ is very uncertain as
the best-fit scale radius lies beyond the radius at which mass measurement
can be made and is therefore strongly dependent on the outermost datapoint of
the mass profile.
The value for $r_{\rm 200}$ is shown for interest, but is not used for any
subsequent analysis.
We instead adopt $r_{\rm 2500}$, which is the most reliable estimate since
it lies well within the radius at which mass measurements are
obtained. We indeed find an excellent agreement between the values of
$r_{\rm 2500}$, $M_{\rm 2500}$ and $f_{\rm gas, 2500}$ obtained from the
NFW fit
to the mass estimated through the $\beta$-model and deprojection methods.
In particular, in the following discussion we adopt the values obtained
with the  $\beta$-model method as their determination is more precise
(cf. Tables \ref{rdelta-beta.tab} and \ref{rdelta-deproj.tab}).

\begin{figure}
\epsscale{.80}
\plotone{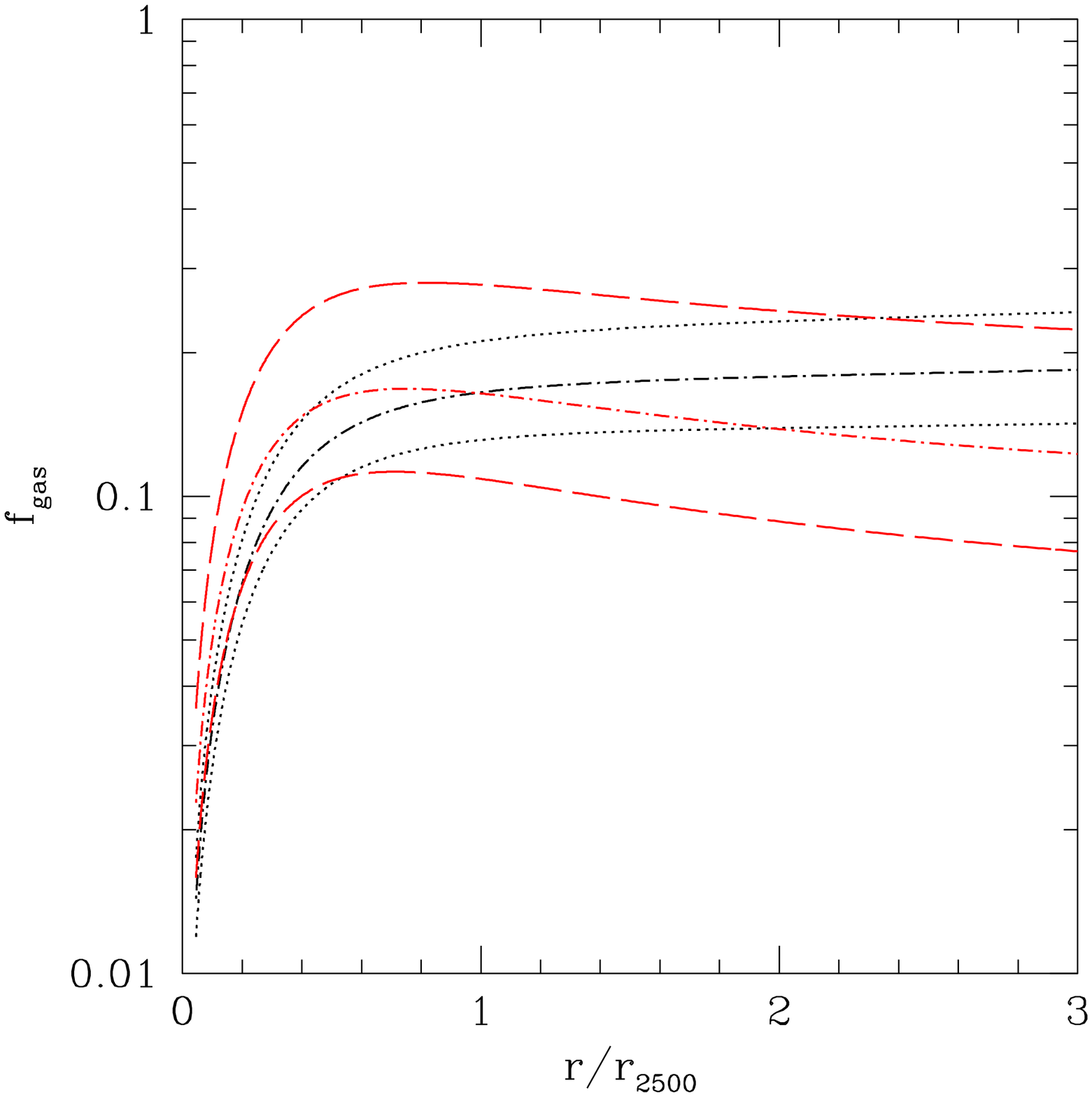}
\caption{
$f_{\rm gas}$ profile with the radial axis scaled in units of $r_{2500}$.
The line colors and styles are the same as in Fig. \ref{nfw.fig}.
\label{fgas-scaled.fig}}
\end{figure}

The gas mass fraction profile scaled in radial units of $r_{2500}$
is shown in Fig. \ref{fgas-scaled.fig}. 
The apparent flattening of $f_{\rm gas}$ outside $r_{2500}$ is in agreement 
with previously observed profiles (Allen et al. 2002, Voigt \& Fabian 2006), 
although it is systematically slightly above them.
In particular, we find $f_{\rm gas, 2500} =0.165 \pm 0.040$,
which is higher than the average value derived in a number of previous
measurements with
\textit{Chandra} (e.g., Allen et al. 2002, Vikhlinin et al. 2006).
The high central gas fraction is close to the global
baryon fraction in the Universe, constrained by CMB observations to be
$\Omega_b / \Omega_m = 0.175 \pm 0.023$ (Readhead et al. 2004, Spergel
et al. 2003).
However, our estimate of the central $f_{\rm gas}$ could be affected
by several effects.
In particular, the high measured central gas mass fraction could result from
an underestimation of the total mass,
supporting the presence of non-thermal pressure in the core.
It could also result from an overestimate of the gas density due to the
boost in emissivity produced by the cavities (see Sect. \ref{l-t.sec}).

\subsection{Scaled temperature profile}
\label{scaled-t.sec}

In order to make a fair comparison between the physical properties
and investigate the scaling relations
it is important to correct for the effects of the central
cooling flow when measuring the characteristic temperature of the cluster.
The average emission-weighted cluster temperature 
is calculated by fitting with a {\ttfamily mekal} model
the spectrum extracted up to the outer radius detected by our X-ray 
observation (6 arcmin), after excising the cooling region (30 arcsec).
We find a value $<T_{\rm X}> = 4.79 \pm 0.13$ keV.

\begin{figure}
\epsscale{.80}
\plotone{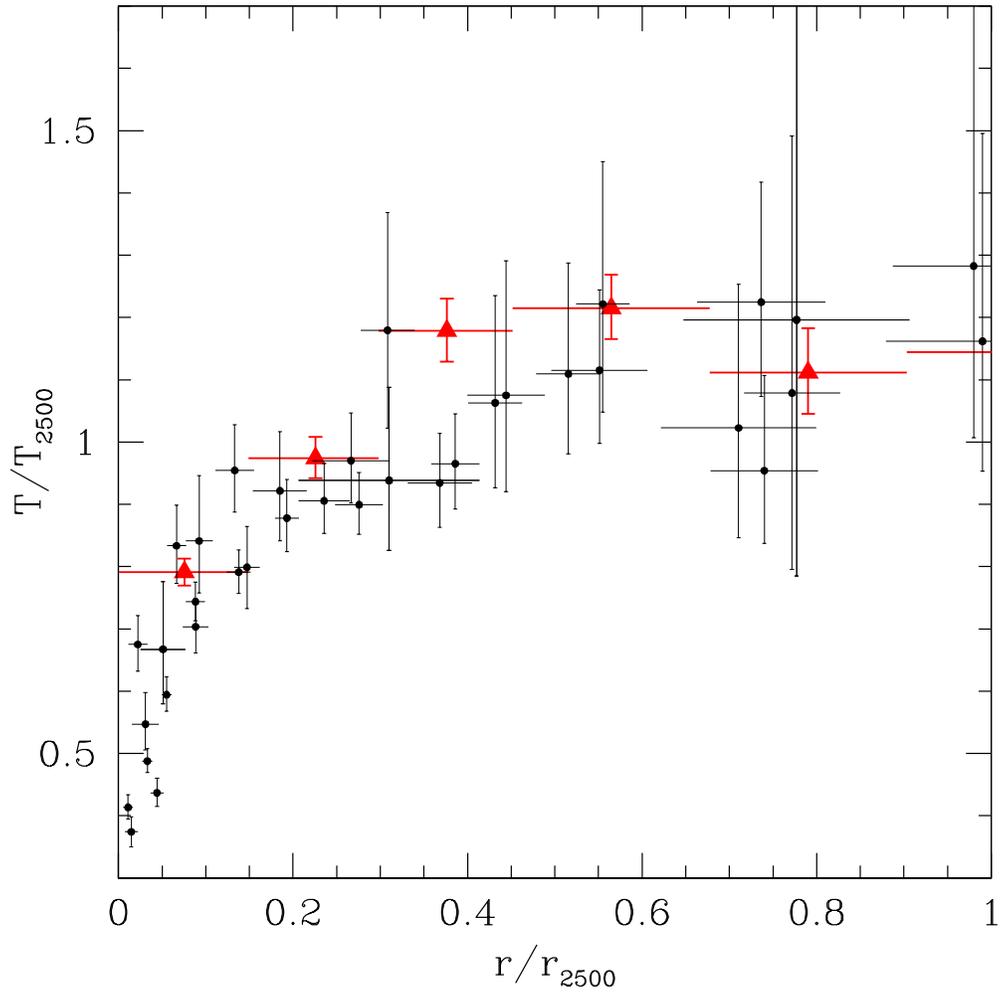}
\caption{
Temperature profile measured for MS0735 (red triangles) overlaid
onto the temperature profile observed for a sample of 6 relaxed clusters
presented by Allen et al. (2001).
The profiles for all clusters are projected and scaled in units of
$T_{2500}$ and $r_{2500}$. 
$T_{2500}$ is estimated within $r_{2500}$ and results 
$T_{2500} = 4.65 \pm 0.08$ keV for MS0735.
\label{temp-scaled-allen.fig}}
\end{figure}

\begin{figure}
\epsscale{.80}
\plotone{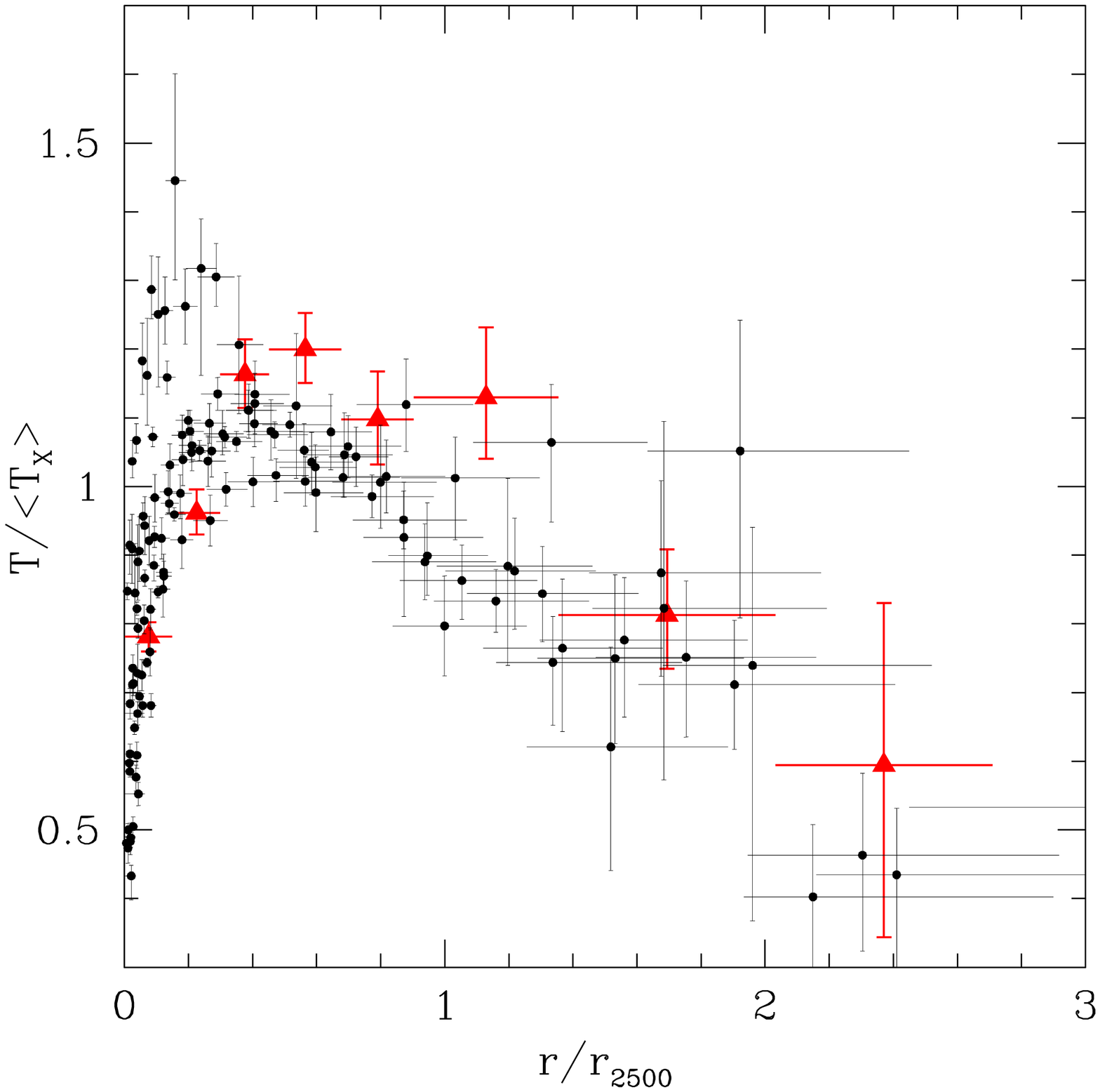}
\caption{
Temperature profile measured for MS0735 (red triangles) overlaid onto the
temperature profile observed for a sample of 12 relaxed clusters presented
by Vikhlinin et al. (2005).
The profiles for all clusters are projected and scaled in radial units of
$r_{2500}$. The temperatures are scaled to the cluster emission-weighted
temperature excluding the central 70 kpc regions.
\label{temp-scaled-alexey.fig}}
\end{figure}

We show in Fig. \ref{temp-scaled-allen.fig} the projected
temperature profile of MS0735 (red triangles) measured in the region
internal to $r_{2500}$. The profile is scaled at an overdensity of
2500 and overlaid to the scaled temperature profiles of a sample of 6
relaxed clusters observed with \textit{Chandra} (Allen et al. 2001).
In Fig. \ref{temp-scaled-alexey.fig} we present the same kind of comparison
with a sample of 12 relaxed clusters for which the temperature profile
has been measured farther out, up to more than 2 $r_{2500}$,
revealing a clear general temperature decline in the outer regions
(Vikhlinin et al. 2005). In this case the temperatures are scaled to the
emission-weighted cluster temperature, excluding the central 70 kpc
region usually affected by radiative cooling.
To be fully consistent with the scaled profiles presented by Vikhlinin et
al. (2005), we thus estimate the emission-weighted temperature used for 
the scaling by extracting the global spectrum of MS0735 in the annular region 
20 arcsec - 6 arcmin. 
We find a value ($kT = 4.71 \pm 0.11$ keV) slightly lower than the ``true''
average emission-weighted temperature, as expected since the effect of cooling 
flow is not completely corrected (the cooling radius in MS0735 is 
$\sim 100$ kpc). 
The overall temperature profile measured for MS0735
is consistent within the scatter of the profiles observed in relaxed cluster,
although for $r>0.5 \, r_{2500}$ its values tend to
lie at the top of the distribution. 

These results indicate that the energetic outburst in MS0735 
does not cause dramatic instantaneous departures from the average 
properties of the ICM, as it has not had a large 
impact on the large-scale temperature profile.  
Of the roughly 30 clusters from the {\it Chandra} archive showing evidence of
AGN outbursts (Rafferty et al. 2006), three have outburts of
comparable energy, MS0735 (McNamara et al. 2005), Hercules A (Nulsen
et al. 2005a) and Hydra A (Nulsen et al. 2005b; Wise et al. 2006). 
These three large outbursts all have ages of $\sim10^8$ yrs.  Dunn et al. 
(2005) find that $\sim70\%$ of cooling flow clusters
currently show signs of outbursts, implying that outbursts are active
most of the time.  If the large outbursts were confined to only these
few clusters, their incidence and ages imply that the clusters have
undergone tens of outbursts since they were formed.  In that case, the
total energy added to the ICM outside the cooling region should have
had a marked effect, contradicting our findings for MS0735.  Thus, the
relatively high incidence of large outbursts is more likely to be due
to them occurring $\sim10\%$ of the time in a significant proportion of
all cooling flow clusters, rather than occurring most of the time in
$\sim10\%$ of clusters.

\subsection{Scaled metallicity profile}
\label{scaled-m.sec}

\begin{figure}
\epsscale{.80}
\plotone{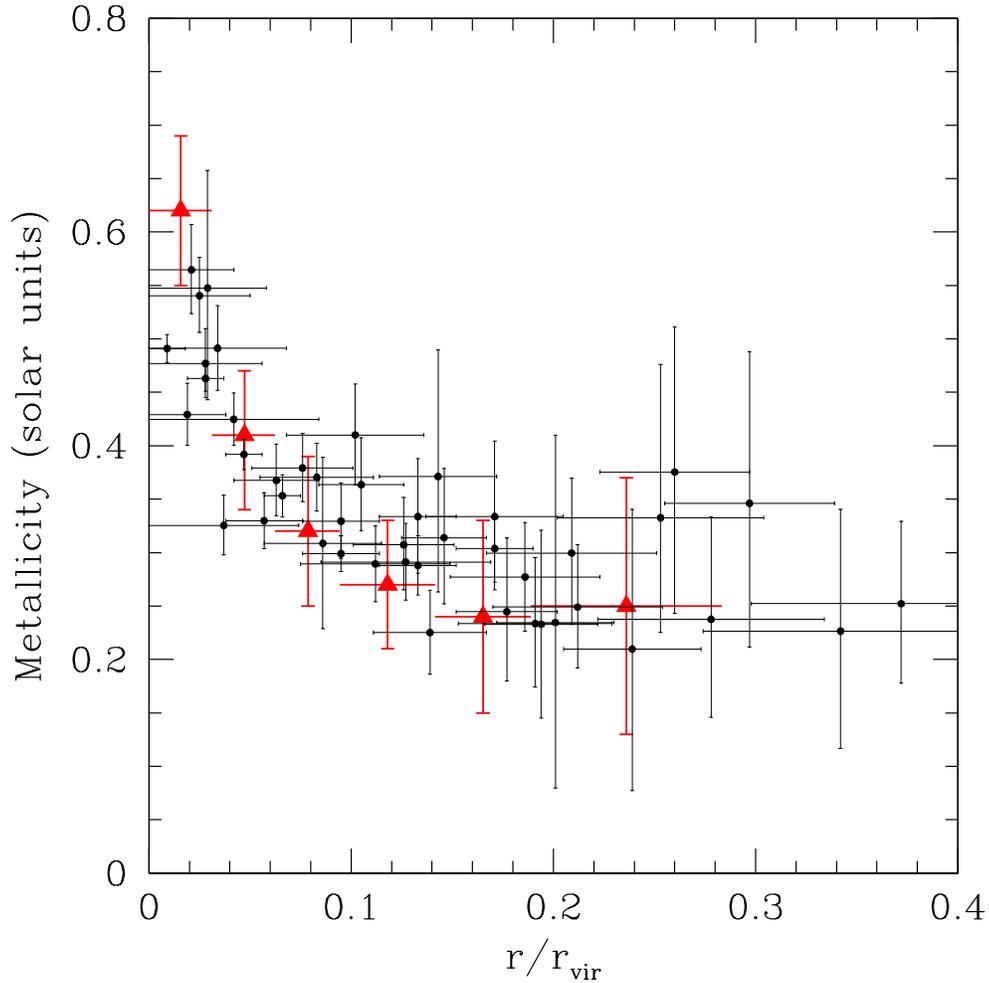}
\caption{
Metallicity profile measured for MS0735 (red triangles) overlaid onto the 
metallicity profile observed for a sample of 9 cooling flow clusters 
presented by De Grandi \& Molendi (2001). 
The profiles for all clusters are projected and scaled in radial units
of $r_{\rm vir}$. The virial radius is estimated from the relation
$r_{\rm vir} = 3.95 \, {\rm Mpc} \, \sqrt{<T_{\rm X}>/ \, 10 \, {\rm keV}}$ 
(Evrard et al. 1996; see also Sect. \ref{r-t.sec}).
$H_0 = 50 \mbox{ km s}^{-1} \mbox{ Mpc}^{-1}$,
$\Omega = 1$, $\Lambda=0$ is assumed.
\label{met-scaled.fig}}
\end{figure}

We studied the metallicity profile in order to investigate the possible
effect of uplifting and outflows on the metal distribution.
Fig. \ref{met-scaled.fig} shows the projected metallicity profile of 9
cooling flow clusters observed with \textit{BeppoSAX},
rescaled to the virial radius (De Grandi \& Molendi 2001).
A strong enhancement in the abundance is found in the central regions.
Overlaid is the metallicity profile that we measure for MS0735
(red triangles). We note that it is fully consistent with the strong central
enhancement in the abundance exhibited by cooling flow clusters,
indicating that the outburst has not smoothed the metal gradient.

How to interpret this result in the context of current models is unclear.
Simulations of metal evolution due to mixing and uplifting by 
buoyant bubbles indicate that low-power, subsonic flows have a 
relatively weak impact on the metallicity gradients found in clusters
(Br\"uggen 2002, Omma et al. 2004), whereas 
powerful outbursts with short active time
can flatten the central metallicity gradients
if the metals are not replenished quickly (Heath et al. 2006).
The average jet power of the outburst in MS0735, as estimated from the 
shock model, is $1.7 \times 10^{46}$ erg s$^{-1}$ (McNamara et al. 2005)
and the observed steep central metallicity profile 
is consistent with the findings of Heath et al. (2006) 
for jet simulations of comparable power.
However, the cavities in MS0735 would increase the jet power significantly,
in which case the dredge up by the rising bubbles is expected to have a 
significant effect in flattening the metallicity gradient.
If such outbursts are rare and
we assume no continuous central metal injection, the fact
that we observe a normal gradient could thus be interpreted as a lack of 
strong evidence for significant metal uplifting or mixing by the cavities
in MS0735.
If instead as discussed above all clusters go through a similar phase, 
the observed agreement would not be surprising as the large outbursts would 
contribute to create the global cluster properties.
Note however that the existing models do not allow us to exclude the 
possibility of some dredging occurring in MS0735.
Indeed, the predictions of numerical simulations depend strongly on the
details of the jet (like opening angle, initial Mach number, duration), 
which can significantly affect the extent of mixing. 
A direct comparison with the metallicity profile observed in MS0735 would
therefore require a tailoring of such simulations to the particular conditions 
of the outburst in this cluster.
Furthermore, observations of broader metallicity profiles compared to the
cD light profiles in cD clusters (De Grandi \& Molendi 2001, David et al. 2001,
Rebusco et al. 2005) indicate that some outward diffusion of enriched
gas due to AGN or mergers is occurring.
Powerful outburst as the one in MS0735 are more likely to produce such an 
effect.
What we conclude from our analysis is therefore that the single outburst in 
MS0735 has not appreciably affected the metal profile compared to other objects,
but we cannot exclude the possibility of some mixing.

\begin{figure}
\epsscale{.80}
\plotone{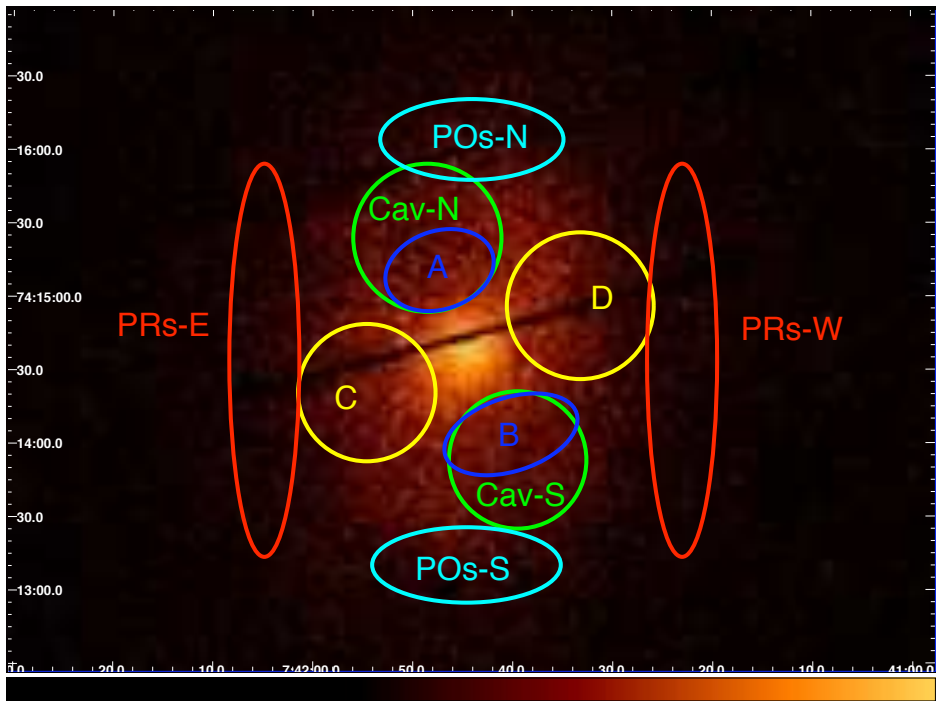}
\caption{
Regions considered to study the abundance distribution
(cf. also the regions in Figs. \ref{cavities.fig} and \ref{shock-a.fig}). 
\label{all-region.fig}}
\end{figure}

\begin{figure}
\epsscale{.80}
\plotone{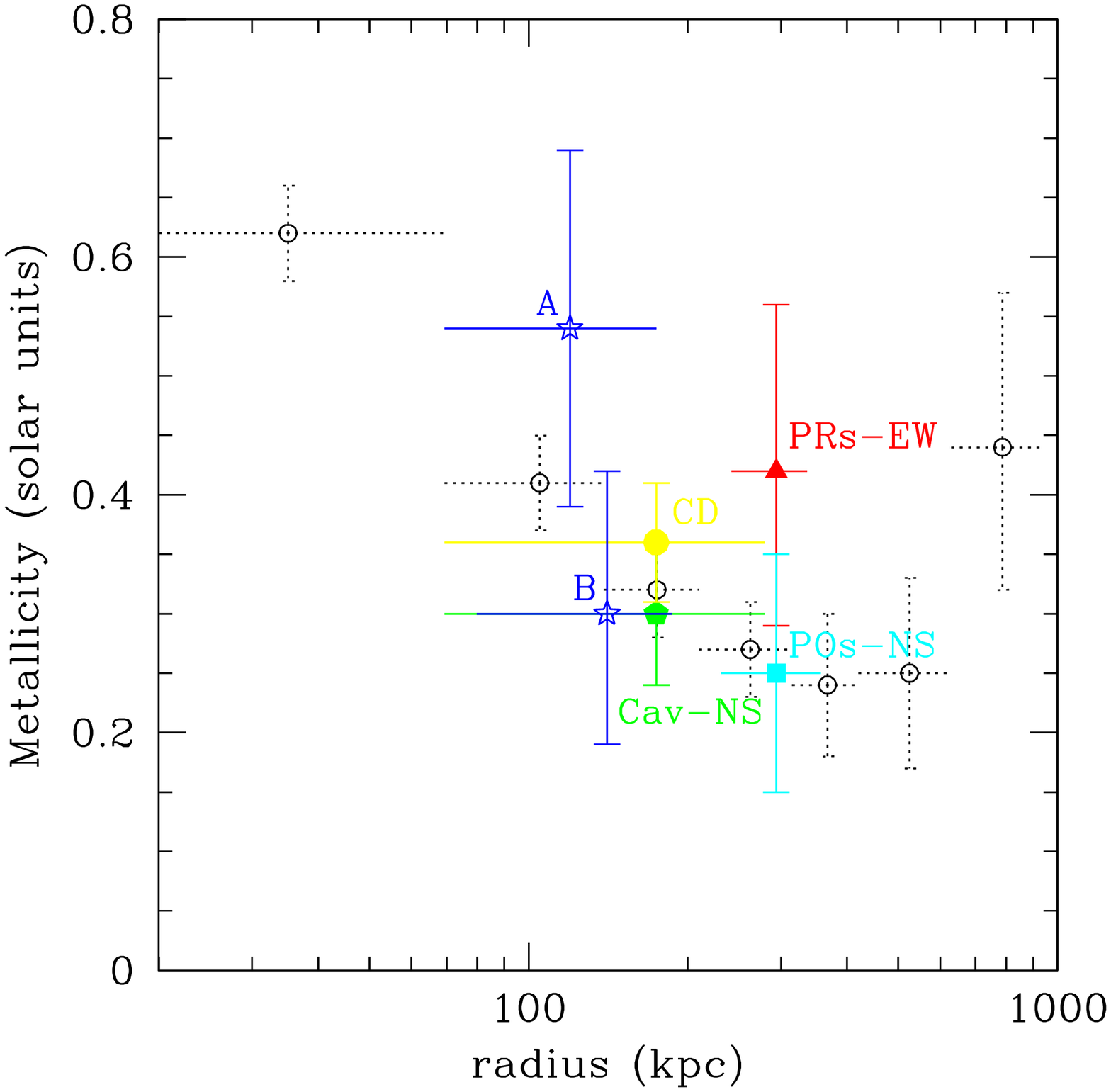}
\caption{
The abundances measured in different regions are indicated with the 
same color scheme used to define the regions in Fig. \ref{all-region.fig}.
For comparison, the azimuthally-averaged metallicity profile is also shown
(open circles).  The error bars are at 1$\sigma$ level.
\label{met-region.fig}}
\end{figure}

On the other hand, we note that when dealing with azimuthally-averaged 
profiles we cannot study the local distribution of metals.
In order to further investigate the metallicity distribution we
extract the spectra in the regions indicated in Fig. \ref{all-region.fig}
and measure the abundance by fitting them with a {\ttfamily mekal} model. 
The results are reported in Table \ref{met.tab} and shown in Fig. 
\ref{met-region.fig}, where we also plot for comparison the 
azimuthally-averaged profile.
Simulations involving uplift by the cavities show that material is entrained 
at the upper (or lower) surface of the rising bubble and in its wake 
(e.g., Churazov et al. 2001), 
and the resulting metal distribution is very elongated along the direction 
of the bubbles (Roediger et al. 2006).
We note that the possibility of making an accurate comparison between
the pattern of abundance that we observe and the prediction of these models
is complicated by the poor photon statistics, which do not allow us to place
definite constraints on the metal distribution. 
Detailed abundance maps of the central cluster regions are 
required for this purpose.

\subsection{$r$ - $T$ relation}
\label{r-t.sec}

A tight size-temperature relationship Is expected, as both of these
quantities reflect the depth of the gravitational potential well in a
virialized halo.
Self-similarity predicts that $r_{\Delta} \propto \sqrt{<T_{\rm X}>}$
(Mohr \& Evrard 1997).
Studies of ensembles of simulated clusters have confirmed this proportionality,
providing a value for the normalization in the relation 
(e.g., Evrard et al. 1996).
Recent \textit{XMM-Newton} observations of a sample of six relaxed galaxy 
clusters hotter than 3.5 keV
find for an overdensity $\Delta = 2500$ that (Arnaud et al. 2005):
\begin{equation}
h(z) \; r_{2500} = (500 \pm 5) \,  {\rm kpc} \; 
\left( \frac{<T_{\rm X}>}{5 \, {\rm keV}} \right)^{0.50 \pm 0.03}  
\label{r2500.eq}
\end{equation}
where the factor $h(z)$ corrects for the evolution expected in the
standard self-similar model (see Eq. \ref{hz.eq}).
In the case of MS0735 the scaling factor is small, being $h(z) = 1.11$
at $z=0.216$. 
This relation allows us to estimate a value of $r_{2500} \sim 440 \pm 5$ kpc, 
which is consistent with the value
of $r_{2500} = 463 \pm 65$ kpc determined from the overdensity profile
(see Sect. \ref{rdelta.sec}).
We therefore find no evidence that the outburst is visibly affecting
the size-temperature relationship.

We note that different studies on samples of galaxy clusters find that
the $r_{\Delta}$-$T$ relation, as deduced from simulated clusters
(Evrard et al. 1996), deviates systematically from the measured values
of $r_{\Delta}$, as inferred from the overdensity profile (Sanderson et al.
2003; Piffaretti et al. 2005). The largest discrepancy is found for
$\Delta = 200$ and in the smallest haloes.
Our result, derived for $\Delta = 2500$ in a massive cluster,
does not contradict these findings. 
The estimate of $r_{2500}$ is more reliable than that of $r_{200}$ as it does
not require much extrapolation of the universal NFW profile 
(see Sect. \ref{rdelta.sec}). 
Furthermore, the impact of additional additional, non-gravitational heating 
is minimal in massive clusters. 
Its effect is most pronounced in poor, cool clusters,
where the extra energy required to account for their observed properties
is comparable to their thermal energy 
(e.g., Ponman et al. 1996; Tozzi \& Norman 2001).


\subsection{$M$ - $T$ relation}
\label{m-t.sec}

Theoretical predictions based on the simplistic assumption of an isothermal
sphere for both the gas represented by its temperature and the collisionless
dark matter particles (e.g. Kaiser 1986; Bryan \& Norman 1998) infer
a self-similar scaling relation between $M_{\rm tot}$ and $T$ at a
given overdensity in the form $M \propto T^{3/2}$.
Various observational studies have found different and sometime conflicting
results regarding the slope and normalization of the $M$-$T$ relation
(e.g., Allen et al. 2001; Finoguenov et al. 2001; Ettori et al. 2002;
Sanderson et al. 2003, Arnaud et al. 2005 and references therein).
The relation derived by Arnaud et al. (2005) for a sub-sample of
six relaxed clusters hotter than 3.5 keV observed with \textit{XMM-Newton}
is consistent with the standard self-similar expectation, 
following the relation:
\begin{equation}
h(z) M_{2500} = (1.79 \pm 0.06)\times 10^{14} {\rm M}_{\odot} 
\left( \frac{<T_{\rm X}>}{5 \, {\rm keV}} \right)^{1.51 \pm 0.11} 
\label{m2500.eq}
\end{equation}
This result is in agreement with \textit{Chandra} observations 
(Allen et al. 2001).
In the case of MS0735, Eq. \ref{m2500.eq} turns into an estimate of 
$M_{\rm 2500} = (1.51 \pm 0.05) \times 10^{14} M_{\odot}$.
By considering the whole \textit{XMM-Newton} sample (ten clusters in the
temperature range [2-9] keV), the relation steepens with a slope 
$\sim 1.70$ (Arnaud et al. 2005) indicating a breaking of self-similarity. 
In this case we estimate 
$M_{\rm 2500} = (1.41 \pm 0.05) \times 10^{14} M_{\odot}$.
Although in better agreement with the $M$-$T$ relation predicted from
the cluster scaling laws, the mass estimate of
$M_{2500} = (1.74 \pm 0.42) \times 10^{14} M_{\odot}$
as derived from the overdensity profile (see Sect. \ref{rdelta.sec})
is still consistent with a steeper slope of the relation considering the
large errors that we measure.


\subsection{$L$ - $T$ relation and pre-heating}
\label{l-t.sec}

The observed relation between X-ray luminosity and gas
temperature in clusters is steeper than expected if cluster growth
were governed by gravity alone.
This steepening is best explained by the addition of heat to
the ICM, and it has been estimated that the excess energies
required to reconcile models and observations lie in the range of
1 to 3 keV per particle (Wu et al. 2000).
The additional non gravitational heating is thought to have
been injected into the gas during an early epoch of star formation
and AGN activity, and therefore it has been called "pre-heating".
Despite this term, there are now strong indications that
powerful AGN outbursts occurring at late times may contribute a
significant fraction of the extra energy 
(McNamara et al. 2005; Nulsen et al. 2005a, 2005b).
In particular, in the case of MS0735 the driving energy of the shock as
determined using a spherical hydrodynamic model is
$E_{\rm s} \approx 5.7 \times 10^{61}$ erg (McNamara et al. 2005).
The AGN outburst in this cluster is therefore heating the gas mass
within 1 Mpc $(\sim 7.7 \times 10^{13} M_{\odot})$ at the level of about
1/4 keV per particle.
The heating level increases to $\sim 0.6$ keV per particle when considering
the gas mass within $r_{\rm 2500}$ $(M_{\rm gas, 2500} \sim 2.9 \times
10^{13} M_{\odot})$.
This is a substantial fraction of the 1-3 keV per particle of excess
energy required to preheat the cluster.

In order to place MS0735 in the observed luminosity vs.
temperature relation for galaxy clusters we estimate that its
X-ray luminosity in the 0.1-2.4 keV energy range is 
$L_X \sim 5.0 \times 10^{44}$ erg s$^{-1}$. 
For a total, emission-weighted temperature of $kT = 4.4$ keV 
the observed luminosity is a factor
$\sim$ 2.6 
higher than that predicted by Markevitch's relation (Markevitch 1998).
The departure of MS0735 from the mean $L$-$T$
relation is reduced when we correct both temperature and
luminosity for the effect of the cooling flow, in the usual manner 
for the study of the scaling relations. 
In order to be fully consistent with the results presented by 
Markevitch (1998),
we estimate the effect of the cooling flow by adopting the
same approach of this author. The corrected temperature is
obtained by averaging the projected temperature 
profile after excluding the
coolest component in the central 20 arcsec bin 
(as estimated from a multiphase spectral fit),
resulting $kT \sim 5.1$ keV.
The corrected luminosity is estimated from the flux observed by masking the 
central 20 arcsec
and then multiplying by a factor 1.06 which accounts for the flux of the 
hot plasma component inside the masked region.  
The corrected luminosity in the 0.1-2.4 keV energy range is 
$L_X \sim 4.2 \times 10^{44}$ erg s$^{-1}$ 
and the corrected bolometric luminosity is 
$L_{\rm bol} \sim 9.0 \times 10^{44}$ erg s$^{-1}$, 
which correspond to a factor $\sim 2$ higher than expected from the 
corrected average temperature of the cluster. 

Note that the cooling flow region in MS0735 is bigger 
(see Sect. \ref{cf.sec}) than the average value adopted by Markevitch
(1998) for the cluster sample, therefore the luminosities and temperature
estimated above are not completely corrected for the effect of cooling flow. 
A more precise correction is obtained by estimating the luminosities from
the spectra extracted after excluding the central 30 arcsec and then adding
back in the luminosity expected from a $\beta$-model profile inside the
masked region. In this case we find values of $L_{[0.1-2.4]} \sim 3.8 \times
10^{44}$ erg s$^{-1}$ and $L_{\rm bol} = 8.1 \times 10^{44}$ erg s$^{-1}$. 
Similarly as done above, the temperature is estimated 
by averaging the temperature profile after excluding the
coolest component in the central 30 arcsec bin, resulting
$kT \sim 5.4$ keV. 
By considering these values ``fully corrected'' for the effect of cooling
flow, the departure from the observed $L$-$T$ relation is slightly
reduced to a factor $\sim 1.7$. 
The corrected luminosity vs. temperature relations for luminosities 
estimated in the X-ray and bolometric bands are plotted in
Figs. \ref{lt-corr.fig} and  \ref{lt-corr-bol.fig}, respectively.
The MS0735 representative points are shown as red full triangles when 
corrected consistently with the method adopted by Markevitch (1998), and as 
blue open triangles when fully corrected for the effect of cooling flow.
The results on temperatures and luminosities estimated in this section 
are summarized in Table \ref{l-t.tab}.

\subsubsection{The ``cavity effect''}
\label{cavity-effect.sec}

\begin{figure}
\epsscale{.80}
\plotone{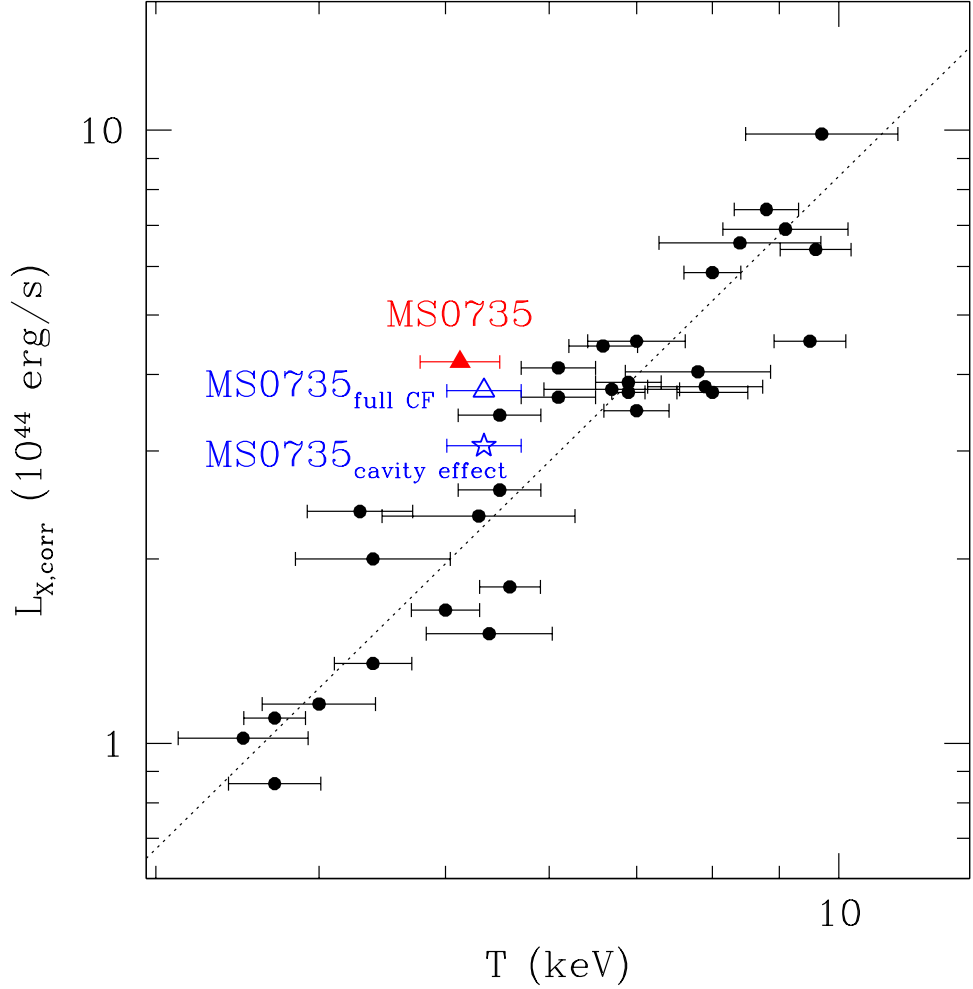}
\caption{
Luminosities in the 0.1-2.4 keV band 
corrected for the effect of cooling flow in the central $\sim$ 70 kpc
vs. emission-weighted temperatures derived excluding
cooling flow components, from Markevitch (1998).
The red full triangle represents MS0735 data from the present observations
corrected consistently with the method adopted by Markevitch (1998).
The blue open triangle represents the MS0735 data fully corrected for
the effect of cooling flow (the cooling flow region in MS0735 is bigger than
the average value adopted by Markevitch (1998) for the cluster sample). 
We also show for interest the data corrected for the cavity effect 
(blue open star).
See text for details.
\label{lt-corr.fig}}
\end{figure}

\begin{figure}
\epsscale{.80}
\plotone{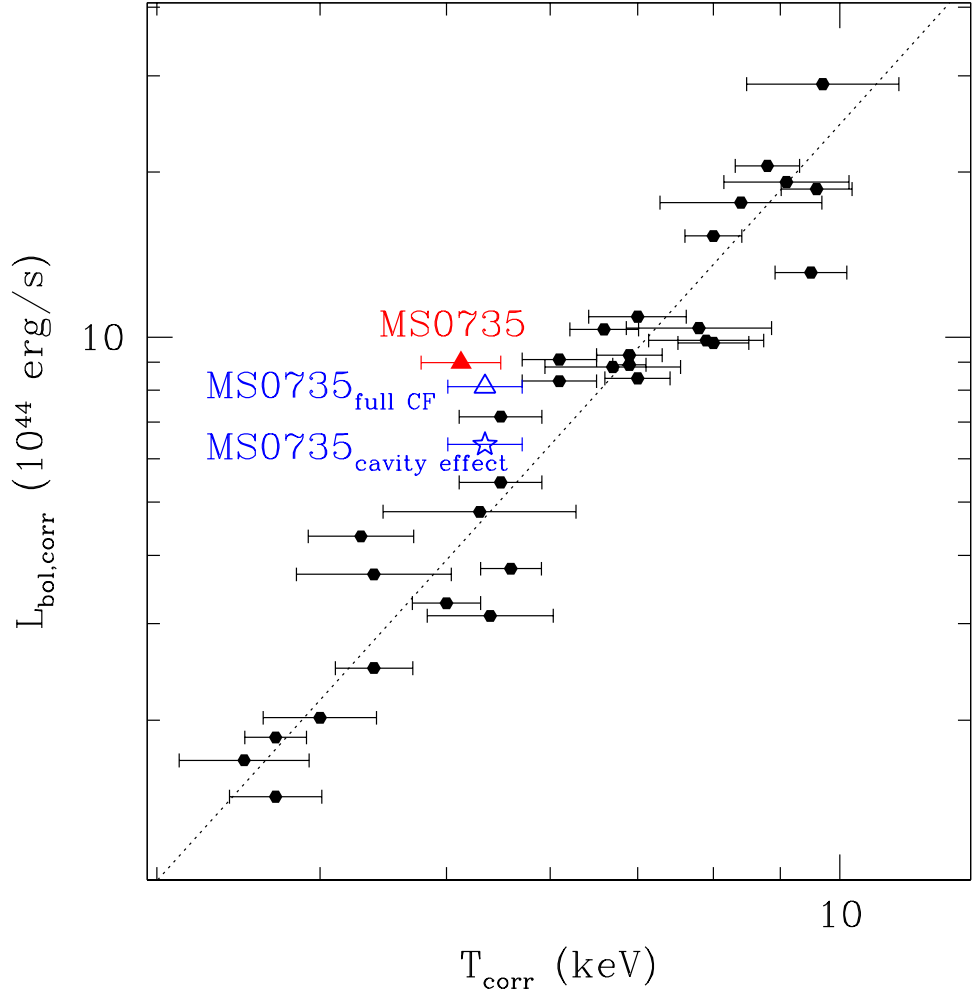}
\caption{
Same as Fig. \ref{lt-corr.fig} for corrected bolometric luminosities. 
\label{lt-corr-bol.fig}}
\end{figure}

\begin{figure}
\epsscale{0.80}
\plotone{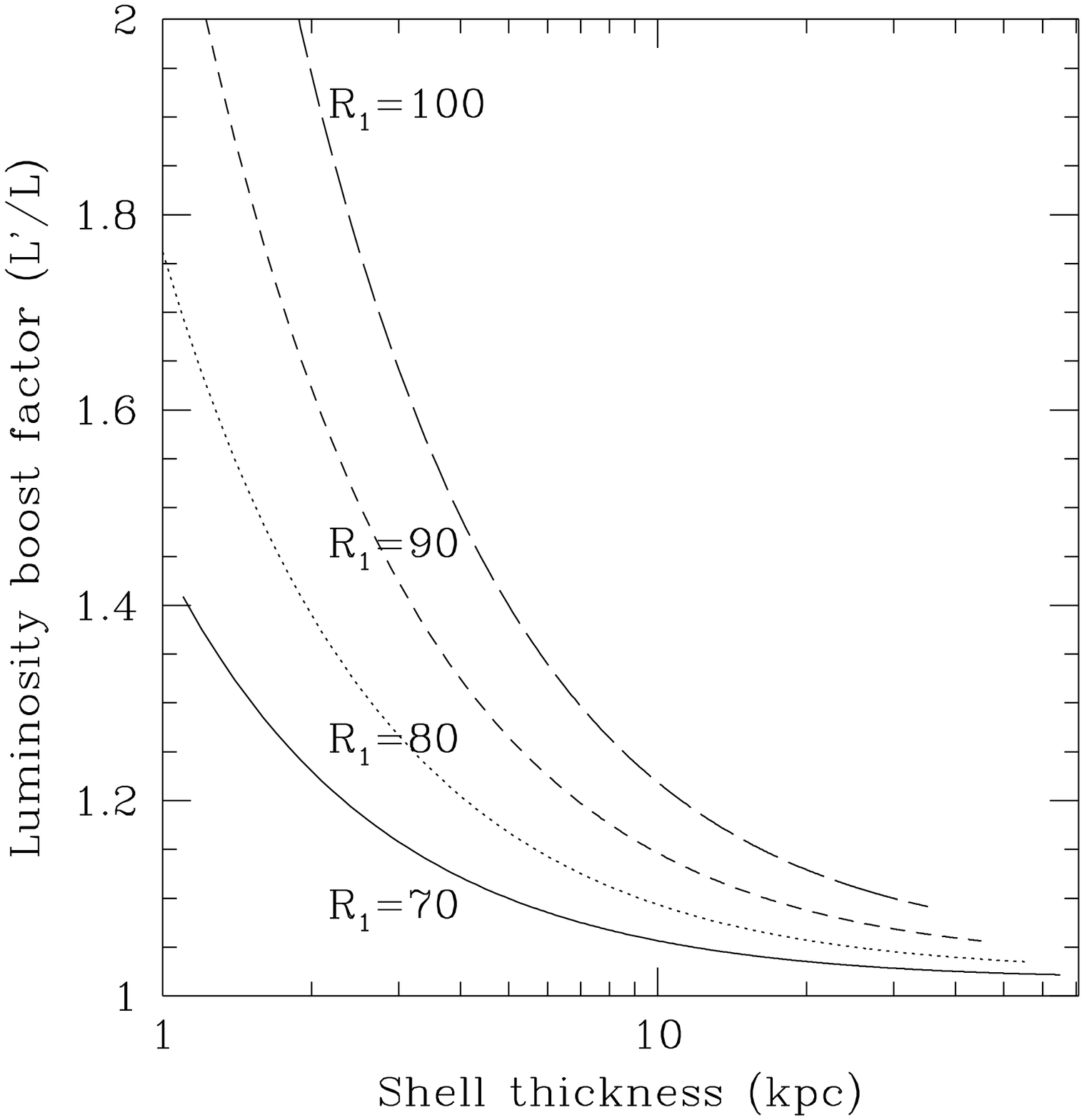}  
\caption{
Estimated luminosity boost factor due to cavity espansion and gas 
compression in the shells, as a function of the thickness of the shell.
Different curves refer to different radius $R_1$ of the cavities 
(units in kpc). 
In this simplified phenomenological model all the gas filling the cavities 
is assumed to be compressed into the bright shells.
For an adiabatic index $\gamma = 5/3$, the maximum compression in a normal 
shock is a factor $f = 4$, so that the ratio of the shell thickness to the 
cavity radius is $R_2/R_1 - 1 > (1 + 1/f)^{1/3} - 1 \simeq 0.077$.
For the cavities observed in MS0735 ($R_1 \sim 100$ kpc), the shell thickness
is $\sim 8$ kpc, leading to a luminosity boost factor \ltsim 25\%.
}
\label{L-boost.fig}       
\end{figure}

The energetic outburst and the consequently
rising cavities uplift the central cool, low-entropy gas up to
large radii, and the same time the compression in the shells
increases the ICM density. This 
effect of cooling flow, cavities and bright shells (that for simplicity
in the following we refer to as the 'cavity effect')
results in an increase of emissivity and thus luminosity. 
By considering a simple phenomenological model of the gas emissivity 
which assumes that all the gas filling the cavities 
is compressed into the bright shells due to the cavity expansion,
we estimate that the luminosity is boosted by a factor
which depends upon the cavity radius and shell thickness. 
For the particular configuration of the cavities observed in MS0735, we
expect an increase in luminosity by a factor as high as about
25\% (see Fig. \ref{L-boost.fig}), consistent with our measurements. 
Note that since the cavities lie
outside the cooling region, their effect is not taken into account 
by the methods adopted above to correct for the cooling flow. 
The emission from the cavity region should also 
be masked before estimating the average properties 
of the ``undisturbed'' cluster.
We thus extract the spectra in the annular region 1.3 - 6 arcmin
(i.e. we excise the central 1.3 arcmin region which includes the cavities).
From these spectra we estimate: 
$L_{[0.1-2.4]} \sim 3.1 \times 10^{44}$ erg s$^{-1}$, 
$L_{\rm bol} = 6.4 \times 10^{44}$ erg s$^{-1}$,
where the missing luminosity expected from a $\beta$-model 
profile inside the central masked region is added back in.
A comparison of these values with those obtained by fully correcting
for the effect of cooling flow allows us to evaluate the impact of 
the cavity effect on the luminosity (cf. blue open triangles and stars in 
Figs. \ref{lt-corr.fig} and  \ref{lt-corr-bol.fig}).
The observed luminosity increase is consistent with that expected by our
simplified model of the structure and emissivity of the gas.

We want to stress that the cavity effect may contribute to 
partially explain the upward departure of MS0735 from the mean 
$L$-$T$ relation, accounting for \ltsim 25\% of the increased luminosity.  
It may also be relevant for the measurements of the gas mass fraction 
in galaxy clusters: the
cavities create lumps in the ICM and this results in an increase
of emissivity and therefore in an overestimate of the gas density.
We find indeed some evidence that the gas mass fraction in MS0735
might be higher than typically observed (see Sect. \ref{rdelta.sec}).

\section{Conclusions}
\label{conclusion.sec}

The main conclusions of this work can be summarized as follows:

\begin{itemize}

\item
The energetic outburst in MS0735 can heat the ICM up to a
level of about 1/4 - 1/2 keV per particle, depending on its
radius of influence, thus contributing a
substantial fraction of the 1 to 3 keV per particle of excess
energy required to preheat the cluster (Sect. \ref{l-t.sec}). 
Only a few outbursts of this magnitude erupting over the life of a 
cluster would be required to preheat it.

\item
MS0735 is a factor $\sim$2 more luminous than expected from its average 
temperature on the basis of the observed $L$-$T$ relation for galaxy
clusters (Sect. \ref{l-t.sec}).
Flux limited samples of distant X-ray clusters may be biased in favor of
detecting clusters with energetic outbursts.

\item
The ICM compression in the bright shells due to the cavity expansion produces 
a boost of the central luminosity of a factor that we estimate should 
be \ltsim 25\%, consistent with our measurements 
(Sect. \ref{l-t.sec}).
This 'cavity effect' may contribute to 
partially explain the upward departure of MS0735 from the mean 
$L$-$T$ relation, and may
also lead to an overestimate of the gas mass fraction 
in qualitative agreement with the high value of 
$f_{\rm gas, 2500}$ that we measure for MS0735 (Sect. \ref{rdelta.sec}). 

\item
The observed scaled temperature and metallicity profiles
are in general agreement with those observed in relaxed clusters 
 (Sects. \ref{scaled-t.sec} and \ref{scaled-m.sec}). 
Also, the quantities we measure for MS0735 are consistent with the 
$r$-$T$ and $M$-$T$ relations predicted by the cluster scaling laws 
(Sects. \ref{r-t.sec} and \ref{m-t.sec}).
This probably indicates that 
violent outbursts such as the one in MS0735 do not cause gross 
instantaneous departures from the average cluster profiles and 
cluster scaling relations (other than the $L$-$T$ relation).  
However, if they are relatively common they may play a role in shaping these
global properties.

\end{itemize}

It is nevertheless hard to draw general conclusions from only one 
object, and therefore in the future we plan to observe and study more 
supercavity systems.


\acknowledgments

We thank the anonymous referee for helpful comments that improved the paper.
We thank S. Ettori for providing the software required to produce the
X-ray colour map in Fig. \ref{tmap.fig}.
We thank S. Allen, A. Vikhlinin, and S. De Grandi for sending the data
used to make the plots in Figs. \ref{temp-scaled-allen.fig},
\ref{temp-scaled-alexey.fig} and \ref{met-scaled.fig}.
M.G. thanks F. Brighenti and B. Maughan for useful comments on the original 
manuscript.
This research is supported by Grant NNG056K87G from NASA's Goddard Space Flight
Center and by NASA Long Term Space Astrophysics Grant NAG4-11025.



\clearpage

\begin{deluxetable}{cccc}
\tablewidth{0pt}
\tablecaption{Spectral fitting in concentric annular regions
\label{profile_proj.tab}
}
\tablehead{
\colhead{Radius} & \colhead{$kT$} &  \colhead{$Z$} & \colhead{$\chi^2$/dof}\\
\colhead{(kpc)} & \colhead{(keV)} & \colhead{($Z_{\odot}$)} & }
\startdata
0-70 (0-$20''$) & $3.68^{+0.10}_{-0.10}$ & $0.62^{+0.07}_{-0.07}$
& 694/641
\\
70-140 ($20''$-$40''$) & $4.53^{+0.16}_{-0.15}$ &
$0.41^{+0.06}_{-0.07}$ & 695/652
\\
140-210 ($40''$-$1'.0$) & $5.48^{+0.24}_{-0.23}$ &
$0.32^{+0.07}_{-0.07}$ & 715/617
\\
210-315 ($1'$-$1'.5$) & $5.65^{+0.25}_{-0.23}$ &
$0.27^{+0.06}_{-0.06}$ & 718/690
\\
315-420 ($1'.5$-$2'$) & $5.17^{+0.33}_{-0.31}$ &
$0.24^{+0.09}_{-0.09}$ & 484/505
\\
420-630 ($2'$-$3'$) & $5.32^{+0.48}_{-0.42}$ &
$0.25^{+0.12}_{-0.12}$ & 582/599
\\
630-945 ($3'$-$4'.5$) & $3.83^{+0.45}_{-0.37}$ &
$0.44^{+0.21}_{-0.18}$ & 858/739
\\
945-1260 ($4'.5$-$6'$) & $2.80^{+1.33}_{-1.18}$ &
$1.11^{+3.37}_{-0.94}$ & 1129/810
\enddata
\tablecomments{Results of the spectral fitting in concentric annular
regions in the [0.4-10.0] keV energy range obtained by fixing the
absorbing column density to the Galactic value ($N_{\rm H} = 3.49
\times 10^{20} {\rm cm}^{-2} $). 
The temperature (in keV) and metallicity 
(in fraction of the solar value, Anders \& Grevesse 1989)
are left as free parameters. Error bars are at the 90\% confidence
levels on a single parameter of interest.}
\end{deluxetable}


\clearpage

\begin{deluxetable}{c|cccc|cccc}
\tablewidth{0pt}
\rotate
\tablecaption{Deprojection analysis
\label{deprojection.tab}
}
\tablehead{
 & \multicolumn{4}{|c|}{MOS1 + MOS2} & \multicolumn{4}{|c}{pn}\\
Radius & \colhead{$kT$} &  \colhead{$Z$} & \colhead{norm} & 
$n_{\rm e}$  & \colhead{$kT$} &  \colhead{$Z$} & \colhead{norm} & 
\colhead{$n_{\rm e}$} \\
(kpc) & \colhead{(keV)} & \colhead{($Z_{\odot}$)} & 
\colhead{($\times 10^{-4}$)} & (cm$^{-3}$) & \colhead{(keV)} & 
\colhead{($Z_{\odot}$)} & \colhead{($\times 10^{-4}$)} & \colhead{(cm$^{-3}$)}
}
\startdata
0-70 & $3.65^{+0.20}_{-0.19}$ & $0.80^{+0.16}_{-0.15}$ &
$7.37^{+0.36}_{-0.36}$ & $0.0210^{+0.0046}_{-0.0046}$  &
$3.20^{+0.18}_{-0.18}$ & $0.59^{+0.15}_{-0.12}$ &
$8.45^{+0.42}_{-0.36}$ & $0.0225^{+0.0050}_{-0.0046}$
\\
70-140 & $3.57^{+0.33}_{-0.28}$ & $0.77^{+0.23}_{-0.20}$ &
$6.64^{+0.48}_{-0.49}$ & $0.0075^{+0.0020}_{-0.0020}$   &
$4.19^{+0.43}_{-0.39}$ & $0.33^{+0.16}_{-0.14}$ &
$9.55^{+0.41}_{-0.58}$ & $0.0090^{+0.0019}_{-0.0022}$
\\
140-210 & $6.24^{+1.03}_{-0.78}$ & $0.21^{+0.19}_{-0.19}$ &
$8.86^{+0.47}_{-0.45}$ & $0.0053^{+0.0012}_{-0.0012}$ &
$4.86^{+0.79}_{-0.63}$ & $0.47^{+0.23}_{-0.22}$ &
$7.55^{+0.55}_{-0.48}$ & $0.0049^{+0.0013}_{-0.0012}$
\\
210-315 & $6.10^{+0.68}_{-0.51}$ & $0.37^{+0.16}_{-0.15}$ &
$12.87^{+0.53}_{-0.55}$ & $0.0035^{+0.0007}_{-0.0007}$ &
$5.80^{+0.83}_{-0.71}$ & $0.22^{+0.16}_{-0.14}$ &
$11.63^{+0.53}_{-0.49}$ & $0.0033^{+0.0007}_{-0.0007}$
\\
315-420 & $5.00^{+0.75}_{-0.57}$ & $0.21^{+0.20}_{-0.17}$ &
$8.65^{+0.71}_{-0.53}$ & $0.0020^{+0.0006}_{-0.0005}$ &
$4.90^{+0.89}_{-0.68}$ & $0.25^{+0.18}_{-0.17}$ &
$8.66^{+0.53}_{-0.48}$ & $0.0020^{+0.0005}_{-0.0005}$
\\
420-630 & $6.00^{+1.05}_{-0.85}$ & $0.30^{+0.26}_{-0.23}$ &
$7.12^{+0.49}_{-0.48}$ & $0.0009^{+0.0002}_{-0.0002}$ &
$6.37^{+1.84}_{-1.37}$ & $0.12^{+0.29}_{-0.12}$ &
$6.23^{+0.37}_{-0.48}$ & $0.0009^{+0.0002}_{-0.0002}$
\\
630-945 & $3.56^{+0.52}_{-0.43}$ & $0.50^{+0.33}_{-0.26}$ &
$5.29^{+0.61}_{-0.59}$ & $0.0004^{+0.0002}_{-0.0001}$ &
$4.83^{+1.70}_{-1.13}$ & $0.15^{+0.37}_{-0.15}$ &
$5.34^{+0.54}_{-0.55}$ & $0.0004^{+0.0001}_{-0.0001}$
\\
945-1260 & -- & -- & -- & -- & $2.71^{+1.70}_{-1.01}$ &
$1.59^{+5.99}_{-1.21}$ & $0.98^{+0.64}_{-0.66}$ &
$0.0001^{+0.0001}_{-0.0001}$
\enddata
\tablecomments{Results of the deprojection
analysis on annular MOS and pn spectra using the XSPEC {\ttfamily
projct$\times$tbabs$\times$mekal} model. The column density is
fixed to the Galactic value and the normalizations are in units of
$10^{-14} n_{\rm e} n_{\rm p} V / 4 \pi [D_{\rm A} (1+z)]^2$. The
fits give $\chi^2$/dof = 3160/2610 and 2786/2635 for MOS and pn,
respectively.}
\end{deluxetable}


\clearpage

\begin{deluxetable}{ccccc}
\tablewidth{0pt}
\tablecaption{Cooling flow analysis
\label{cf.tab}
}
\tablehead{
\colhead{Par.} & \colhead{Mod. A} &  \colhead{Mod. B} & \colhead{Mod. C}
& \colhead{Mod. D}
}
\startdata
$kT$ & $3.87^{+0.09}_{-0.09}$ &  $7.61^{+0.45}_{-1.28}$   & $6.07^{+1.27}_{-0.63}$ 
& $4.25^{+0.15}_{-0.18}$  
\\
$Z$ & $0.57^{+0.05}_{-0.05}$ &  $0.52^{+0.06}_{-0.05}$  & $0.52^{+0.05}_{-0.06}$ 
& $0.59^{+0.05}_{-0.05}$ 
\\
Norm & $1.7\times 10^{-3}$ & $7.9 \times 10^{-5}$  &  $9.1 \times 10^{-4}$ 
&  $1.4 \times 10^{-3}$
\\
$kT_{\rm low}$ &  --- &  $1.45^{+0.23}_{-0.14}$   & $2.27^{+0.41}_{-0.39}$ 
&   $0.1$
\\
Norm$_{\rm low}$   & --- & ${\dot M}=260$ &  $6.7 \times 10^{-4}$ 
& ${\dot M}=40$
\\
$\chi^2$/dof & 891/787 &  839/785  & 839/785 
& 861/786
\enddata
\tablecomments{
The best-fit parameter values and 90\% confidence limits of the spectral
analysis in the central 30 arcsec region.
Temperatures ($kT$) are in keV, metallicities ($Z$) as a fraction of the solar
value
and normalizations in units of $10^{-14} n_{\rm e} n_{\rm p} V / 4 \pi
[D_{\rm A} (1+z)]^2$ as done in XSPEC (for the {\ttfamily mkcflow} model the
normalization is parameterized in terms of the mass deposition rate $\dot{M}$,
in $\mbox{M}_{\sun} \mbox{ yr}^{-1} $). See text for details.
}
\end{deluxetable}


\clearpage

\begin{deluxetable}{c|ccc}
\tablewidth{0pt}
\tablecaption{Cavity analysis
\label{cavity.tab}
}
\tablehead{
 & \colhead{Par.} & \colhead{{\ttfamily mekal}} & 
\colhead{{\ttfamily mekal+mekal}}
}
\startdata
         & $kT_1$       & $5.24^{+0.36}_{-0.33}$ &  $13.1^{+27.6}_{-4.7}$  \\
         & $Z$          & $0.31^{+0.12}_{-0.11}$ &  $0.28^{+0.13}_{-0.12}$  \\
Cavity N & Norm$_1$     & $4.9\times 10^{-4}$    & $2.3 \times 10^{-4}$  \\
         & $kT_2$       &  ---                   &  $3.45^{+0.96}_{-1.00}$  \\
         & Norm$_2$     & ---                    & $ 2.7 \times 10^{-4}$ \\
         & $\chi^2$/DOF & 394/313                &  385/311  \\
\tableline
         &  $kT_1$      & $5.00^{+0.36}_{-0.33}$ &  $0.43^{+0.21}_{-0.15}$   \\
         &  $Z$         & $0.28^{+0.12}_{-0.11}$ &  $0.28^{+0.14}_{-0.10}$ \\
Cavity S &  Norm$_1$    & $3.9\times 10^{-4}$    &  ---  \\
         & $kT_2$       &  ---                   &  $5.31^{+0.41}_{-0.36}$ \\
         & Norm$_2$     & ---                    &  $ 3.8 \times 10^{-4}$ \\
         & $\chi^2$/DOF & 288/287                &  275/285  \\
\enddata
\tablecomments{
The best-fit parameter values and 90\%
confidence limits of the spectral analysis in the cavity region.
Temperatures ($kT$) are in keV, metallicities ($Z$) as a fraction
of the solar value and normalizations in units of $10^{-14} n_{\rm
e} n_{\rm p} V / 4 \pi [D_{\rm A} (1+z)]^2$ as done in XSPEC.
}
\end{deluxetable}


\clearpage

\begin{deluxetable}{ccccc}
\tablewidth{0pt}
\tablecaption{Shock analysis
\label{shock.tab}
}
\tablehead{
\colhead{Region} & \colhead{source counts}& \colhead{$kT$}& \colhead{Z}& 
\colhead{$\chi^2$/dof}\\
 & \colhead{(MOS+pn)} & \colhead{(keV)} & \colhead{$(Z_{\odot})$} & 
}
\startdata
{\bf PRs-E } & 3081 & $6.15^{+1.25}_{-0.88}$ &
$0.34^{+0.33}_{-0.29}$ & 143/122
\\
{\bf PRs-W } & 2361 & $4.82^{+0.71}_{-0.56}$ &
$0.44^{+0.36}_{-0.30}$ &  87/100
\\
{\bf POs-N } & 2958 & $5.29^{+0.72}_{-0.62}$ &
$0.27^{+0.26}_{-0.22}$ & 116/110
\\
{\bf POs-S } & 3122& $5.57^{+0.94}_{-0.74}$ &
$0.25^{+0.26}_{-0.24}$ & 113/118
\\
{\bf (PRs-E) $+$ (PRs-W)} & 5442 & $5.34^{+0.58}_{-0.50}$ &
$0.42^{+0.23}_{-0.20}$ & 301/227
\\
{\bf (POs-N) $+$ (POs-S) } & 6080 & $5.43^{+0.60}_{-0.46}$ &
$0.25^{+0.17}_{-0.16}$ & 233/233
\\
{\bf PRs} (annulus) & 13187 & $5.24^{+0.36}_{-0.33}$ &
$0.21^{+0.10}_{-0.09}$ & 440/461
\\
{\bf POs} (annulus) & 22655 & $5.63^{+0.26}_{-0.24}$ &
$0.29^{+0.07}_{-0.07}$ & 647/634
\enddata
\tablecomments{
Results from the spectral fitting in the regions
indicated in Figs. \ref{shock-a.fig} and \ref{shock-b.fig}. 
The fit is performed in the [0.4-10.0] keV energy range by fixing the 
absorbing column density to the Galactic value. 
When fitting simultaneously two different regions at the same radial distance 
from the center (e.g., PRs-E + PRs-W), the normalizations of each camera are 
linked in order to have the same value in the two regions.
Error bars are at the 90\% confidence levels on a single parameter of 
interest.
}
\end{deluxetable}


\clearpage

\begin{deluxetable}{cccc}
\tablewidth{0pt}
\tablecaption{Results from $\beta$-model analysis
\label{rdelta-beta.tab}
}
\tablehead{
\colhead{$\Delta$} & \colhead{$r_{\Delta}$} & \colhead{$M_{\rm tot}$} & 
\colhead{$f_{\rm gas}$}     \\
    & \colhead{(kpc)}    & \colhead{($10^{14} M_{\odot}$)} &  
}
\startdata
   200   &  1718 (218)  & 7.12 (1.96)  & 0.188 (0.052) \\
   500   &  1107 (145)  & 4.77 (1.26)  & 0.181 (0.048) \\
  1000   &  775 (105)   & 3.27 (0.83)  & 0.176 (0.046) \\
  2500   &  463 (65)    & 1.74 (0.42)  & 0.165 (0.040) \\
\enddata
\tablecomments{Characteristic radii $r_{\Delta}$, total mass $M_{\rm tot}$ and
mass gas fraction $f_{\rm gas}$
for various overdensities $\Delta$ derived from the NFW fit to the mass
estimated through the $\beta$-model.
All the quantities are estimated within $r_{\Delta}$
(1$\sigma$ errors in parentheses).
}
\end{deluxetable}


\clearpage

\begin{deluxetable}{cccc}
\tablewidth{0pt}
\tablecaption{Results from deprojection analysis
\label{rdelta-deproj.tab}
}
\tablehead{
\colhead{$\Delta$} & \colhead{$r_{\Delta}$} & \colhead{$M_{\rm tot}$} & 
\colhead{$f_{\rm gas}$}     \\
    & \colhead{(kpc)}    & \colhead{($10^{14} M_{\odot}$)} &  
}
\startdata
   200   &  2230 (650)  & 15.6 (8.78)  & 0.109 (0.061) \\
   500   &  1340 (410)  & 8.46 (4.42)  & 0.124 (0.065) \\
  1000   &  875 (280)   & 4.71 (2.33)  & 0.141 (0.070) \\
  2500   &  465 (158)   & 1.76 (0.81)  & 0.165 (0.076) \\
\enddata
\tablecomments{
Same as Table \ref{rdelta-beta.tab} for various overdensities $\Delta$
derived from the NFW fit to the mass estimated through the deprojection
method.
}
\end{deluxetable}


\clearpage

\begin{deluxetable}{ccc}
\tablewidth{0pt}
\tablecaption{Measured abundances
\label{met.tab}
}
\tablehead{
\colhead{Region} & \colhead{Z} & \colhead{$\chi^2$/dof}\\
 & \colhead{$(Z_{\odot})$} &   
}
\startdata
{\bf PRs-E } & $0.34^{+0.33}_{-0.29}$ & 143/122
\\
{\bf PRs-W } & $0.44^{+0.36}_{-0.30}$ &  87/100
\\
{\bf POs-N } & $0.27^{+0.26}_{-0.22}$ & 116/110
\\
{\bf POs-S } & $0.25^{+0.26}_{-0.24}$ & 113/118
\\
{\bf Cav-N } & $0.31^{+0.12}_{-0.11}$ & 394/313
\\
{\bf Cav-S } & $0.28^{+0.12}_{-0.11}$ & 288/287
\\
{\bf A } & $0.54^{+0.22}_{-0.20}$ & 188/174
\\
{\bf B } & $0.30^{+0.16}_{-0.15}$ & 173/210
\\
{\bf C } & $0.36^{+0.09}_{-0.10}$ & 346/380
\\
{\bf D } & $0.37^{+0.11}_{-0.11}$ & 371/335
\\
{\bf (PRs-E) $+$ (PRs-W)} & $0.42^{+0.23}_{-0.20}$ & 301/227
\\
{\bf (POs-N) $+$ (POs-S) } & $0.25^{+0.17}_{-0.16}$ & 233/233
\\
{\bf (Cav-N) $+$ (Cav-S) } & $0.30^{+0.08}_{-0.08}$ & 869/605
\\
{\bf A $+$ B} & $0.40^{+0.13}_{-0.12}$ & 497/389
\\
{\bf C $+$ D} & $0.36^{+0.07}_{-0.07}$ & 902/720
\\
\enddata
\tablecomments{
Abundance measured in the regions
indicated in Fig. \ref{all-region.fig}. 
The {\ttfamily mekal} fit is performed in the [0.4-10.0] keV
energy range by fixing the absorbing column density to the
Galactic value. 
When fitting simultaneously two different regions at the same radial distance 
from the center (e.g., C + D), the normalizations of each camera are 
linked in order to have the same value in the two regions.
Error bars are at the 90\% confidence levels on a
single parameter of interest.
}
\end{deluxetable}


\clearpage

\begin{deluxetable}{llccc}
\tablewidth{0pt}
\tablecaption{Results for $L$-$T$ relation
\label{l-t.tab}
}
\tablehead{
\colhead{Band} & \colhead{Data}  & \colhead{$T$} & \colhead{$L_{\rm obs}$} &
\colhead{$L_{\rm rel}^{\rm M}$}   \\
\colhead{(keV)}  &  & \colhead{(keV)} & \colhead{($10^{44}$ erg/s)} & 
\colhead{($10^{44}$ erg/s)} 
}
\startdata
0.1-2.4    & total                & $ 4.4 \pm 0.1$  & 5.00   & 1.89   \\
0.1-2.4    & CF corr.$^{\rm M}$     & $ 5.1 \pm 0.4$  & 4.19   & 2.07   \\
0.1-2.4    & CF fully corr.       & $ 5.4 \pm 0.4$  & 3.76   & 2.26   \\
0.1-2.4    & cavity effect corr.  &      ---        & 3.06   & ---    \\ 
Bolom.     & CF corr.$^{\rm M}$     & $ 5.1 \pm 0.4$  & 8.98   & 4.20   \\
Bolom.     & CF fully corr.       & $ 5.4 \pm 0.4$  & 8.12   & 4.69   \\
Bolom.     & cavity effect corr.  &      ---        & 6.37   & ---    \\
\enddata
\tablecomments{
Temperatures and luminosities measured for MS0735. 
The effect of the cooling flow (CF) has been evaluated 
by employing two different approaches.
{\it CF corr.$^{\rm M}$} indicates the correction adopted by Markevitch (1998):
the temperature is obtained by averaging the temperature profile after 
excluding the coolest component in the central 20 arcsec bin 
(as estimated from a multiphase spectral fit);
the luminosities are estimated from the flux observed by masking the 
central 20 arcsec and then multiplying by a factor 1.06 which accounts for 
the flux of the hot plasma component inside the masked region.  
{\it CF fully corr.} indicates a more precise correction which takes into
account the ``true'' extent of the cooling region in MS0735: 
the temperature is estimated by averaging the temperature profile after 
excluding the coolest component in the central 30 arcsec bin;
the luminosities are estimated from the spectra extracted after excluding the 
central 30 arcsec and then adding back in the luminosity expected from a 
$\beta$-model profile inside the masked region. 
$L_{\rm rel}^{M}$ indicates the luminosity expected from 
the luminosity vs. temperature relation (Markevitch 1998). 
The luminosities corrected for the cavity effect are also reported.
See text (Sect. \ref{cavity-effect.sec}) for details.
}
\end{deluxetable}


\end{document}